\documentclass[twocolumn]{aastex631}

\usepackage{amsmath,mathtools}

\newcommand\omc{$\omega$\,Cen}
\newcommand\newtext{}

\begin{document}

\title{oMEGACat IV: Constraining Ages of Omega Centauri sub-giant branch stars with \textit{HST} and MUSE} 

\correspondingauthor{Callie Clontz}
\email{clontz@mpia.de}

\author[0009-0005-8057-0031]{C. Clontz}
\affiliation{Department of Physics and Astronomy, University of Utah, Salt Lake City, UT 84112, USA}
\affiliation{Max Planck Institute for Astronomy, K\"onigstuhl 17, D-69117 Heidelberg, Germany}

\author[0000-0003-0248-5470]{A. C. Seth}
\affiliation{Department of Physics and Astronomy, University of Utah, Salt Lake City, UT 84112, USA}

\author[0000-0002-4442-5700]{A. Dotter}
\affiliation{Department of Physics and Astronomy, Dartmouth College, Hanover, NH 03755 US}

\author[0000-0002-5844-4443]{M. H\"aberle}
\affiliation{Max Planck Institute for Astronomy, K\"onigstuhl 17, D-69117 Heidelberg, Germany}

\author[0000-0002-2941-4480]{M. S. Nitschai}
\affiliation{Max Planck Institute for Astronomy, K\"onigstuhl 17, D-69117 Heidelberg, Germany}

\author[0000-0002-6922-2598]{N. Neumayer}
\affiliation{Max Planck Institute for Astronomy, K\"onigstuhl 17, D-69117 Heidelberg, Germany}

\author[0000-0002-0160-7221]{A. Feldmeier-Krause}
\affiliation{Department of Astrophysics, University of Vienna, T\"urkenschanzstrasse 17, 1180 Wien, Austria}
\affiliation{Max Planck Institute for Astronomy, K\"onigstuhl 17, D-69117 Heidelberg, Germany}

\author[0000-0002-7547-6180]{M. Latour}
\affiliation{Institut für Astrophysik und Geophysik, Georg-August-Universität Göttingen, Friedrich-Hund-Platz 1, 37077 Göttingen, Germany}

\author[0000-0003-2512-6892]{Z. Wang}
\affiliation{Department of Physics and Astronomy, University of Utah, Salt Lake City, UT 84112, USA}

\author[0000-0001-8052-969X]{S. O. Souza}
\affiliation{Max Planck Institute for Astronomy, K\"onigstuhl 17, D-69117 Heidelberg, Germany}

\author[0000-0002-6072-6669]{N. Kacharov}
\affiliation{Leibniz Institute for Astrophysics (AIP), An der Sternwarte 16, 14482 Potsdam, Germany}

\author[0000-0003-3858-637X]{A.\ Bellini}
\affiliation{Space Telescope Science Institute, 3700 San Martin drive, Baltimore, MD, 21218, USA}

\author[0000-0001-9673-7397]{M. Libralato}
\affiliation{INAF, Osservatorio Astronomico di Padova, Vicolo dell’Osservatorio 5, Padova,I-35122, Italy}

\author[0000-0002-1670-0808]{R. Pechetti}
\affiliation{Astrophysics Research Institute, Liverpool John Moores University, 146 Brownlow Hill, Liverpool, L3 5RF, United Kingdom}

\author[0000-0003-4546-7731]{G. van de Ven}
\affiliation{Department of Astrophysics, University of Vienna, T\"urkenschanzstrasse 17, 1180 Wien, Austria}

\author[0000-0002-1212-2844]{M. Alfaro-Cuello}
\affiliation{Facultad de Ingenier\'{i}a y Arquitectura, Universidad Central de Chile, Av. Francisco de Aguirre 0405, La Serena, Coquimbo, Chile}

\begin{abstract}
 We present age estimates for over 8100 sub-giant branch (SGB) stars in Omega Centauri (\omc) to study its star formation history. Our large data set, which combines multi-wavelength \textit{HST} photometry with MUSE metallicities, provides an unprecedented opportunity to measure individual stellar ages. We do this by fitting each star's photometry and metallicity with theoretical isochrones, that are embedded with an empirical [C+N+O]-[Fe/H] relation specifically for \omc. The bulk of the stars have ages between 13 and 10 Gyr, with the mean stellar age being 12.08 $\pm{0.01}$ Gyrs and the median age uncertainty being 0.68 Gyrs. From these ages we construct the most complete age-metallicity relation (AMR) for \omc\ to date.  We find that the mean age of stars decreases with increasing metallicity and find two distinct streams in the age-metallicity plane, hinting at different star formation pathways. We derive an intrinsic spread in the ages of 0.75$\pm{0.01}$ Gyr for the whole cluster, with the age spread showing a clear increase with metallicity. We verify the robustness of our age estimations by varying isochrone parameters and constraining our systematics. We find the C+N+O relation to be the most critical consideration for constraining the AMR. We also present the SGB chromosome map with age information. In the future, these stellar ages could be combined with chemical abundances to study age differences in subpopulations, and uncover the chemical evolution history of this massive nuclear star cluster.
\end{abstract}


\section{Introduction} \label{sec:intro}

Omega Centauri (\omc) is the most massive star cluster in the Milky Way \citep[$ \rm 3.5 \times 10^{6} M_{\odot}$;][]{Baumgardt_2018}. It is also uniquely complex, with its stars having a nearly 2 dex spread in metallicity \citep{Johnson_2010, Johnson_2020, Nitschai_2024} and a large number of subpopulations \citep[e.g.][]{Bellini_2017c}.  In addition, it is on an unusual, mildly inclined, retrograde orbit in the Milky Way \citep{Dinescu_1999}. 

One proposed explanation for \omc's atypical features is that it is actually the surviving dense core of a tidally stripped galaxy \citep{Norris_1996,Bekki_2003,Hilker_2004,Johnson_2010,Villanova_2014}. Such nuclear star clusters (NSCs) commonly have stellar populations with a wide range of metallicities and extended star formation histories \citep[e.g.][]{Kacharov_2018,Fahrion_2021}.  
The association of \omc\ with other halo stars, thought to be associated with its progenitor galaxy, was proposed by \cite{Majewski_2012}, who found a large number of stars with similar kinematics and also showed common abundance anomalies between these stars and \omc.  With the advent of Gaia, many people have tried associating \omc\ with its progenitor merger event. 
While initial studies suggested a possible association with the Sequoia merger event \citep{Myeong_2019,Forbes_2020}, recent work has strongly suggested that \omc\ is the stripped galaxy nucleus of the Gaia-Enceladus/Sausage merger event based on its orbit, mass, age, and metallicity \citep{Lee_1999, Lee_2009, Pfeffer_2021,Callingham_2022, Limberg_2022}.

The formation of NSCs is an open question of active investigation. Studying \omc's formation history will increase our understanding of many aspects of NSCs, including the ages of stellar populations, mass assembly, dynamical evolution, star formation mechanisms, and the galactic environment.

There have been several works investigating the dominant formation mechanisms for galactic nuclei, (summarized in \citealt{Neumayer_2020}). 
They are thought to begin via the merging of globular clusters, which drift toward the galactic nucleus due to dynamical friction. Later, there is gas infall and subsequent in-situ star formation inside this mixed abundance environment \citep{Neumayer_2020}, creating a complex enrichment and star formation history (SFH) in these objects. There are many more processes that affect the star formation history in a dense galactic center than occur in globular cluster environments, making this scenario appealing for explaining \omc's complexity. 

Some galactic nuclei are thought to become compact star clusters through  tidal-stripping. As a nucleated satellite galaxy falls into its central host (as \omc's previous host did with our Milky Way), tidal forces first remove the outer parts of the galaxy, eventually leaving only the dense core. Simulations have shown that NSCs can survive the disruption process and that in a typical Milky Way mass galaxy we expect several (up to $6$) stripped galaxy nuclei \citep{Pfeffer_2014,Kruijssen_2019}.

\omc\ being the remnant of an accreted nuclear star cluster has significant implications for understanding the assembly history of the Milky Way. The properties of \omc, including its size, mass, orbital characteristics, kinematics, and stellar ages give us insights into the galaxy merger and accretion history that contributed to the growth of our galaxy. Analyzing the chemical abundances and stellar populations within \omc\ can reveal the conditions and chemical makeup of the galaxy from which it originated. Differences in elemental abundances between \omc\ and stars in the Milky Way's disk can help us understand the nature of the interactions of these systems. This information aids in piecing together the complex history of interactions between the Milky Way and its satellite galaxies. 

In addition, we know that the Milky Way has its own dense stellar core, whose assembly mechanisms are likely similar to that of \omc. Therefore, studying \omc\ provides constraints on the evolution of our own galactic nucleus.  

Due to the wide spread in metallicities, which indicates a complex assembly history for \omc\, there have been numerous efforts to measure stellar ages and determine its SFH.
\cite{Hilker_2004} performed isochrone fitting using color, magnitude, and metallicity information for 447 subgiant branch (SGB) and main-sequence turnoff (MSTO) stars in \omc\ and found an age spread of around 3 Gyrs.
\cite{Villanova_2007} measured relative ages for 80 SGB stars and found 4 distinct groups with a large but not fully constrained age spread among them (at least 3 Gyrs). 
\cite{Joo_2013} constructed synthetic CMDs for \omc\ \citep[based on Yonsei-Yale isochrone models,][]{Yi_2001} and found an age spread of $\sim$1.7 Gyrs across the 5 identified subgroups. A follow-up study by \cite{Villanova_2014} examined the spectra of 172 SGB stars and found an age spread of 1.5 to 3 Gyrs is needed to explain the abundance variations seen across the 6 populations they identify. Each of these studies discusses how their assumptions of alpha and helium abundance could affect their age spread determinations. They also stress how important modeling the C+N+O abundance as a function of metallicity is for achieving more precise age and age spread estimates. \cite{Tailo_2016} incorporated the observed C+N+O abundance from \cite{Marino_2012} into their population synthesis model and found that no significant age spread ($<0.5$ Gyrs) is needed to fit the CMD if large helium abundance and high C+N+O enhancement are assumed for the most metal-rich population ($\rm [Fe/H] > -0.8$). 

It is no surprise that the literature provides an inconsistent picture of the intrinsic age spread. \omc's populations are uniquely complex, and large variations are seen not just over metallicity but also in the abundances of stars at a given metallicity \citep{Johnson_2010, Nitschai_2023,Nitschai_2024}.  This poses many challenges to accurate age estimation \citep{Marino_2012,Tailo_2016}.  

Stellar evolution lifetimes and evolutionary tracks can vary significantly based on the chemical composition of the star, and furthermore, determining abundances of individual stars is observationally expensive. 
Additionally, there are several sources of systematic uncertainties, including distance and extinction, that contribute to imprecise absolute ages (though relative ages and age spread measurements are conserved).

Age is a useful measure for tracing the complex evolutionary history of multi-population star clusters. In this paper, we estimate stellar ages in \omc\ for an unprecedented sample of SGB stars ($>$8100).  
These measurements combine high-precision photometry and spectroscopic metallicities together with newly developed, specifically tuned isochrone models to better understand the age spread among its subpopulations. We focus on the SGB region as it minimizes apparent age differences due to varying helium abundance, and we use the photometry measurements with filters having minimal dependencies on light-element variations.

In Section \ref{sec:challenges} we discuss the challenges for constraining ages due to our model assumptions. In Section \ref{sec:data} we describe the data reduction of the MUSE and \textit{HST} observations, in Section \ref{sec:methods} we outline our methods to measure stellar ages, and in Section \ref{sec:stellar_ages} we discuss our results. Section \ref{sec:systematics} discusses constraints on our systematic uncertainties. Discussion and conclusions are presented in Sections \ref{sec:discussion} and \ref{sec:conclusions}. 

\section{Model Choices and the Challenges for Constraining Ages in $\omega$ Cen} \label{sec:challenges}
In this section we highlight the necessary stellar model assumptions and physical parameter choices and each of their contributions to the complexity of measuring absolute ages of individual stars in \omc.  This includes the distance and extinction to the cluster, as well as the C+N+O, [$\alpha$/Fe], and helium abundance variations that are a result of the complex set of subpopulations which comprise \omc.

\subsection{Distance and Reddening Estimates} \label{subsec:dist_extinct_uncertainties}
The distance to \omc\ has been measured through various techniques including eclipsing binaries, RR Lyrae stars, CMD fitting and Gaia parallaxes. However, these methods each have limitations, and therefore the literature values for the distance have a significant spread. 
A review by \cite{Baumgardt_2021} found the mean value of the literature distances to be $\rm 5.426 \pm{0.047} \ kpc$. We adopt this value which gives a distance modulus of $\rm(m - M_0) = 13.672 \pm{0.019}$.

A common method for measuring the distance to a cluster is direct isochrone fitting to the MSTO region with the assumption of foreground extinction. However, such an estimation has uncertainty due to the degeneracy of distance and extinction. 
In this study, we firstly applied the literature values of distance and extinction (E(B-V) = 0.12 from \cite{Harris_1996} (2010 version) to fit isochrones to CMD. However, when using these values the isochrones do not cover their relevant data points on the CMD. According to recent studies on pulsar DM measurements \citep{Zhang_2023} and differential reddening distribution from photometry \citep{Pancino_2024}, there is considerable evidence that \omc\ has higher extinction within half-light radius.

Therefore, we decided to fit for a new reddening value after fixing the distance to the literature value from  \citet{Baumgardt_2021}. Details of this empirical correction, which uses reference clusters with similar metallicity to \omc's dominant population but with very low extinction, are given in Appendix \ref{appendix:empirical_reddening_corr}. We derive a best-fit reddening of E(B-V) = 0.185 (uncertainties discussed in Appendix \ref{appendix:empirical_reddening_corr}).  We also note this value is in rough agreement with foreground reddening estimates based on the Na~D absorption lines of the MUSE spectroscopy (Wang et al., {\em in prep}).  We discuss the effect of distance on our age constraints in Section \ref{subsec:dist_systematics}.
 
\subsection{C+N+O vs. [Fe/H]} \label{subsec:cno_fe_uncertainties}
There are many known correlations and anti-correlations among the light element abundances in globular clusters (including \omc) that result from hydrogen fusion via the CNO cycle, Ne-Na chain, and Mg-Al cycle. These nucleosynthetic processes require increasing star burning temperature and thus occur at different rates within a population \citep{Gratton_2019}. 
Because these reactions only catalyze the hydrogen burning, the total number of heavy elements stays constant, e.g. for the CNO cycle products, the total number of C+N+O atoms is constant. However, unlike in typical globular clusters, the extremely varied stellar populations in \omc show strong variations in C+N+O.

It is known that C+N+O abundance affects the CMD position of both the MSTO and SGB and thus isochrone fitting.  Several studies on the abundances of subpopulations in \omc\ \citep{Marino_2012, Bellini_2017c, Milone_2018, Milone_2020} have shown strong evidence for higher metallicity populations being accompanied by an increase in the total abundance of C+N+O. \cite{Marino_2012} studied the spectra of 77 RGB stars and found that C+N+O increases with increasing metallicity, by about 0.5 dex between [Fe/H] = -2.0 to -0.9. They find that the C+N+O enhancement can cause stars to appear $\sim$1-2 Gyr older, with metal-rich stars being the most affected. We fit the data in the \citet{Marino_2012} paper and used this to create isochrone models that are specific to \omc\ (see Section \ref{subsec:iso_models}). 

\subsection{Helium Abundance Variations} \label{subsec:helium_uncertainties}
Variations in helium abundance are also known to play a major role in isochrone shape and position, with larger differences on the Main Sequence (MS) and RGB, and minimal differences along the SGB \citep[e.g.][]{Milone_2018}. Helium-enhanced stars (and their model isochrones) tend to be bluer than their helium normal counterparts (see central and right panels of Fig.~\ref{fig:sgb_region}). 
Several studies \citep{Norris_2004, Piotto_2005, Sollima_2005, King_2012, Joo_2013} have suggested that the `blue MS' in \omc\ must be helium enhanced to explain why they seem to be more metal-rich than their redder counterparts. 
There have been several studies \citep{Piotto_2005, Tailo_2016, Joo_2013, Milone_2018, Latour_2021} that find it necessary to include helium-enhanced populations to reconstruct \omc's CMD. Other studies compare spectral features to find a range of helium values (up to $\Delta Y=+0.15$ \citealt{King_2012, Reddy_2020}) among the stars.

No existing study has examined the variations in helium abundance with metallicity due to the difficulty of observationally constraining helium in individual stars.  Therefore, we focus on the SGB to minimize the effects of helium on our results.  We also constrain its impact on our derived SGB star ages by recalculating ages with helium-enhanced isochrone models. These results are discussed in Section \ref{subsec:helium_systematics}. 

\subsection{$\alpha$ Abundance Variations} \label{subsec:alpha_uncertainties}
The $\alpha$ elements, such as Si, Ti, Mg, and Ca, are produced by the $\alpha$ process, most commonly occurring in massive stars and supernovae. 
\cite{Johnson_2010} found that the stars of \omc\ tend to have Si, Ca and Ti abundance enhancements (a good proxy for overall alpha abundance) of $ \sim +0.3$, though there is an overall spread of more than 0.5 dex. 
There is also a slight trend in alpha enhancement increasing with metallicity. For this study, we choose to focus on the C+N+O relation with respect to metallicity and therefore set $ \rm{[\alpha/Fe]}$ to a fixed value of 0.3. We discuss the effect of alpha abundance variations in Section \ref{subsec:alpha_systematics}.

\section{Data} \label{sec:data}
The strength of this project comes from compiling new and archival photometric and spectroscopic catalogs to create the most complete picture of \omc\ to date. 

\subsection{MUSE Spectroscopic Metallicities} \label{subsec:metallicities}

The spectroscopic data used for this study were taken with the Multi-Unit Spectroscopic Explorer (MUSE) \citep{Bacon_2010} mounted on the Very Large Telescope (VLT) at the Paranal Observatory in Chile. \newtext{MUSE observations cover the optical range (480-930 nm) with a resolving power that increases with wavelength from 1770 to 3590.} There are 10 multi-epoch pointings of the central regions of \omc\ that were obtained in a Guaranteed Time Observing (GTO) program and provided spectra for approximately 75,000 unique stars. More information is given in \cite{Kamann_2018}, \cite{Husser_2020}, and \cite{Latour_2021}. 
Additional MUSE observations were taken in 2022 as part of program 105.20CG.001 (PI: N. Neumayer) which provided 75 new wide-field-mode (WFM) pointings and 12 new adaptive-optics (AO) pointings covering \omc\ out to the half-light radius (4.65’, 7.04 pc, \citealt{Baumgardt_2018}) and providing spectra for over 300,000 stars. Full details of the catalog can be found in \cite{Nitschai_2023}. 

The \citet{Nitschai_2023} catalog combines the metallicities provided by the GTO team (for over 58,000 stars) with the new extended MUSE mosaic (of over 303,000 stars) to create the largest spectroscopic catalog for any star cluster to date. In both cases \texttt{spexxy}\footnote{https://github.com/thusser/spexxy} is used to analyse the extracted spectra; this code uses full-spectrum fitting to constrain the line-of-sight velocity and derive the overall metallicity [M/H] of individual stars. For the new catalog, over 160,000 stars have reliable metallicity estimates (SNR $>$ 10). For this work, we convert the [M/H] to [Fe/H] via the following equation given by \cite{Salaris_1993}: 

\begin{equation}
    \rm{[Fe/H] = [M/H] - \log(0.638 \times 10^{[\alpha/Fe]} + 0.362)},
\end{equation}
with $ \rm{[\alpha/Fe]}$ set to 0.3 \citep{Johnson_2010}. 
This conversion relies on the detailed abundance of each star and is therefore a source of systematic uncertainties in our results, though found to be on the order of the metallicity error (0.08 dex). Also, this conversion does well in replicating the [Fe/H] distributions measured in other studies \citep{Nitschai_2023,Nitschai_2024}.  We additionally correct our metallicities for atomic diffusion as discussed in detail in \citet{Nitschai_2023}.  

\subsection{\textit{HST} Photometry} \label{subsec:photometry}

This work combines decades of \textit{Hubble Space Telescope (HST)} observations by using the recent catalog of \citet{haeberle_2024a}, \newtext{available at \dataset[https://doi.org/10.17909/26QJ-G090]{https://doi.org/10.17909/26QJ-G090}}.  This catalog includes an analysis of around 800 images from the Advanced Camera for Surveys (ACS) Wide Field Channel and the Wide Field Camera 3 (WFC3) Ultraviolet VISible (UVIS) Channel including numerous calibration images, archival images, and new images taken as part of program 16777 (PI: Seth).  The catalog contains both proper motion measurements and photometry for $\sim$1.4 million stars covering a nearly contiguous spatial region out to the half-light radius.  The photometric coverage is nearly complete in six filters: ACS/WFC results in F435W, F625W, and F658N \citep[previously presented in][]{Anderson_2010}, and WFC3/UVIS results in F275W, F336W, and F814W.  In addition WFC3/UVIS F606W photometry is included in the catalog based on the 184 images taken in this filter (mostly for calibration) at the center of the cluster; these data include stars covering half the radial range as the other filters.

The photometric reduction of the images was performed using {\tt KS2} software (written by Jay Anderson, see \citealt{Bellini_2017a}), and included image-by-image zeropoint corrections and charge transfer efficiency (CTE) corrections.  In addition, spatially varying empirical photometric corrections were made in each filter based on CMD fitting to account for variable extinction and other photometric issues (see Appendix B of \cite{haeberle_2024a} for details).  Errors were estimated based on repeat measurements as a function of instrumental magnitude, and all measurements in each filter were combined using a weighted mean.  We note that these errors do not account for the varying degree of crowding which changes spatially, thus when we have more than 1 measurement, we rescale these errors by the ${\chi_{\rm Reduced}}$ of the combined photometric measurement when the $\chi^2_{\rm Reduced} > 1$.  

For this paper, we primarily use the F606W, F625W and F814W photometry.  CMDs using the F606W-F814W color (or F625W-F814W) have minimal positional differences along the SGB due to light-element abundance variations and helium abundance variations.

After deriving ages, we also use the photometry in the bluer filters (F275W, F336W, and F435W), which are sensitive to light element abundance variations, to analyze the age distribution in the ``chromosome map'' \citep[e.g.][]{Milone_2017a}.  

To account for extinction effect, we use the re-estimated E(B-V)=0.185 mag in Section \ref{subsec:dist_extinct_uncertainties}. Then, we use the conversions from the Padova CMD database\footnote{http://stev.oapd.inaf.it/cgi-bin/cmd\_3.7} to get ${\rm A_{\lambda} / A_{V}}$ values for our \textit{HST} filters:
these are ${\rm A_{\lambda} / A_{V}}$ = 0.90941 for F606W and 0.59845 for F814W, respectively.  Thus, we subtract $ \rm A_{\lambda, 606} = 0.52155$ mag and $ \rm A_{\lambda,625} = 0.34321$ mag to obtain our extinction-corrected photometry.  

\subsection{Data Quality Cuts} \label{subsec:data_quality_cuts}
To select our sample of SGB stars from our combined spectroscopic and photometric catalogs, we make a number of quality cuts that ensure that the stars are both members of \omc, and that their metallicity and photometry measurements are reliable. To ensure we include only cluster members we perform 3-sigma clipping of the proper motion measurements \citep{haeberle_2024a}(corresponding to a velocity within 3 mas\,yr$^{-1}$ (51.44 km\,s$^{-1}$) of the \omc\ system value) and require the line-of-sight velocity membership probability \citep{Nitschai_2023} to be higher than 80\%. For the photometric data, we require all stars to have more than one measurement in both F606W and F814W and a high-quality flag = 1. The quality cuts relating to the MUSE metallicities include 'Flag'=1 which requires a membership probability $>$ 95 \%, SNR $>$ 10, as well as several quality parameters relating to the  \texttt{spexxy} fit and \texttt{pampelmuse} extraction \citep[full details in][]{Nitschai_2023}. We also require the metallicity error fall in the range: $ \rm 0.01 < \sigma_{[Fe/H]} < 0.2$, and that each star has a metallicity in the range of $\rm -2.5 < [Fe/H] < -0.5$ which covers all but three of the SGB stars that meet our other cuts and enables us to focus our isochrone creation on a limited range of metallicities. Within the SGB selection box discussed below, a total of 9,129 stars remain after these quality cuts from an original sample of 13,816 stars with available measurements. The velocity cuts remove 568 stars, the photometric cuts 2,792 stars, and the spectroscopic cuts 1,740 stars, independent of one another. Using the F606W filter for our analysis limits us to the inner 4.5 parsecs of the cluster but greatly improves the precision of our age constraints due to repeat measurements in F606W leading to much lower photometric errors (more discussion in Section \ref{subsec:f625w}).

We show the combined MUSE metallicity measurements and HST photometry for our SGB sample after all quality cuts in Fig.~\ref{fig:sgb_region}. Coloring each star by its metallicity (left panel) makes the discrete sequences more visible.

\section{Methods} \label{sec:methods}

\subsection{Selection of SGB Stars} \label{subsec:sample_selection}

One major challenge for constraining ages of stars with isochrone fitting is that the
helium abundance plays a large role in the location of stars across the full CMD and is hard to spectroscopically constrain the exact values in individual stars. One way around this is to narrow in on the SGB where the color differences between two helium abundance isochrones is minimized \citep{Piotto_2014, Bellini_2017c}. In addition, because the SGB is a short-lived part of a star’s evolution, the CMD location of stars on the SGB provides a more direct and precise measurement of its age. Along the SGB atomic diffusion effects are minimal, and our spectroscopic observations have high Signal-to-Noise resulting in reliable metallicities.
We fit for ages using the F606W $-$ F814W CMD.  These filters have no strong lines from elements that vary in globular cluster multiple populations and minimize differences between helium abundances that are present in wider filter combinations.

\begin{figure*}
\centering
\includegraphics[width = \textwidth]{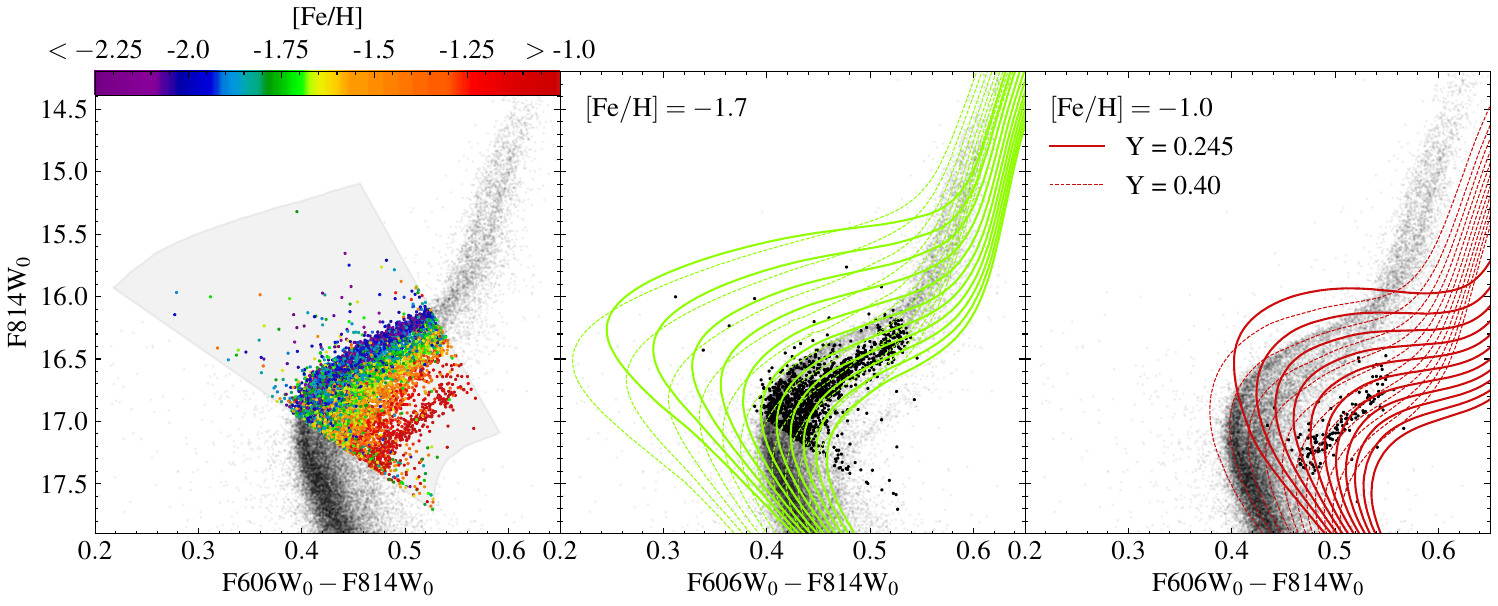}
\caption{\textbf{SGB Selection \& Relevant Isochrones:} The left panel shows our initial SGB region selection colored by the MUSE metallicity. The middle panel shows stars with $-1.75 < {\rm [Fe/H]} < -1.65$ and isochrones (described in Section~\ref{subsec:iso_models}) with [Fe/H] = -1.7  and ages ranging from 5 to 15 Gyr (top to bottom) overplotted in green. The solid lines are the solar scaled helium abundance isochrones and the dashed lines are the Y$=$0.40 models. The right panel is similar to the middle, now with ${\rm [Fe/H]} = -1.0$ and isochrones colored in red. Our selection utilizes the region where the models are the most similar.}
\label{fig:sgb_region}
\end{figure*}

In Fig.~\ref{fig:sgb_region} we show our sample selection on the SGB, colored by metallicity in the left panel. In the central panel we show stars with $-1.75 < \rm[Fe/H] < -1.65$ along with isochrone models (details in Section \ref{subsec:iso_models}) with $\rm[Fe/H] = -1.7$, representing the median metallicity of the cluster. The solid lines are primordial helium isochrones with ages from 5 to 15 Gyrs in 1 Gyr increments. The dashed lines are helium enhanced models which are shown to demonstrate their similarities along the SGB region and how they diverge as they turn on the RGB as well as below the MSTO. The right panel is the same except for stars with $\rm[Fe/H] \sim -1.0$.  We note that the helium-enhanced isochrones do seem to fit the higher metallicity in the right panel stars better than Y$=$0.245 isochrones while the two helium models seem to bracket the median metallicity sample in the central panel.  

We select stars along the SGB in this CMD.  The bounds of our SGB selection (shown in gray in Fig.~\ref{fig:sgb_region}) were determined primarily by where the difference between the Y$=$0.245 and 0.4 isochrone models is on the order of the median combined photometric errors (0.027 mag). The upper and lower bounds are chosen to be within the area covered by the youngest most metal-poor isochrone and the oldest most metal-rich isochrone respectively. We make additional cuts after measuring ages to obtain our final sample (see Section \ref{sec:stellar_ages} for details). Our CMD selection region contains over 9,000 SGB stars, which is the largest combined photometric-spectroscopic dataset for any cluster.

\subsection{Isochrone Models} \label{subsec:iso_models}

To make accurate age estimates it is necessary to have an isochrone model grid that is well-tuned to \omc's stellar populations and is finely sampled in age and metallicity.  

Our isochrone model grids are built from the Dartmouth Stellar Evolution Database \citep{Dotter_2007, Dotter_2008}. To obtain the most accurate ages, we incorporate new data on the C+N+O and $\alpha$ abundance to create a new set of isochrones designed specifically for \omc. The composition is based on the \citet{Grevesse_1998} solar abundance pattern \citep[in keeping with][]{Dotter_2008}. The Dartmouth Stellar Evolution Database models set carbon and nitrogen abundances using a solar-scaled composition and included oxygen with the $\alpha$-capture elements; each of the $\alpha$-capture elements is enhanced in lockstep with the value of [$\alpha$/Fe].

\citet{Marino_2012} found a clear trend of increasing C+N+O abundance with increasing [Fe/H]. While we cannot make direct measurements of the C+N+O abundance from our low-resolution spectra, we can use our [Fe/H] measurements to estimate each star's C+N+O enhancement using the data from \citet{Marino_2012}.

Figure 8 of \citet{Marino_2012} shows a roughly linear trend between $ \rm{\log \epsilon([C+N+O])}$ and [Fe/H]. We perform linear regression on this set of points and arrive at a linear relation: 
\begin{equation}
    \rm{\log \epsilon([C+N+O])  = 1.48 \times [Fe/H] + 10.32}
\end{equation}

\noindent We use this C+N+O vs. [Fe/H] relation to generate a new set of isochrones.  In these new isochrones, we decouple oxygen from the $\alpha$-abundance consideration and fold it in with the C+N+O enhancement based on [Fe/H]. We use these isochrones to perform our primary age constraints and we will refer to this model set as ``C+N+O-relation" hereafter. 

We utilize the [Si/Fe] and [Fe/H] data from \citet{Johnson_2010} to model the relation of the $\alpha$ abundance with metallicity in \omc, finding a fairly constant value of ${\rm [Si/Fe]} = 0.3$ for ${\rm [Fe/H]} \leq -1.50$ and a weak ($\sim$0.1 dex) increase in [Si/Fe]-abundance above this value. With this we chose to assume a constant [$\alpha$/Fe]=0.3 for our primary age constraints, but also create [$\alpha$/Fe]=0.2 isochrones that we use to analyze how big an impact the [$\alpha$/Fe] has on our age determinations.

We have chosen our CMD region and filter combination to limit the impact of the helium abundance on our age determinations.  We therefore do not attempt to constrain the abundance of helium as a function of [Fe/H].  Our default ages assume a primordial helium abundance (Y=0.245+1.5Z), but we also create C+N+O-relation models with enhanced helium (Y=0.4) as shown in Fig.~\ref{fig:sgb_region}; these two isochrones cover the range of expected helium abundances \citep{Reddy_2020}.  We analyze the robustness of our results by comparing differing isochrone sets in Section~\ref{sec:systematics}.  This comparison includes the original Dartmouth Isochrones with a no  C+N+O vs. [Fe/H] trend included, and oxygen folded into the [$\alpha$/Fe] enhancement. This set of isochrones we refer to as ``C+N+O-fixed" hereafter. 

To use our isochrone models for fitting ages, we feed the model grids into two partner codes, \texttt{IsoInterpFeh} and \texttt{IsoSplit} \citep{Dotter_2007}, we create a grid of isochrones with $\rm -2.5 < [Fe/H] < -0.5$ in $0.01$ dex increments and  $\rm 5.0 \leq Age \leq 15.0$ Gyr in steps of $0.25$ Gyr. We chose to sample ages greater than the age of the universe to ensure extended coverage of the Probability Density Function (PDF) for reliable Gaussian fitting. We add the distance modulus to the provided absolute magnitudes in each filter to obtain the extinction corrected apparent magnitudes. Lastly, we interpolate along the path of each isochrone, creating 1,000 points between $14 < \rm{F814W} < 18$ to obtain uniform fine spacing.

\subsection{Age \& Uncertainty Measurements}  \label{subsec:age_age_uncertainty_measurements}
With this new, robust and flexible isochrone grid tuned to \omc's stellar populations, we constrain the ages of individual stars using a maximum likelihood method that combines the CMD position and spectroscopic metallicities. 

We follow closely the methods outlined in \cite{Alfaro-Cuello_2019}. We explore the age-metallicity space of our given isochrones and constrain the posterior probability function using Bayes’ theorem: 

\begin{equation}
\rm{P(\tau | V_{obs},I_{obs},Z_{obs}) \propto P(V_{obs},I_{obs},Z_{obs} | \tau ) \times P(\tau)}
\end{equation}

where $\tau$ is the age of the star, and the normalized likelihood function: $\rm{P(V_{obs},I_{obs},Z_{obs} | \tau)}$ of the observables at a given age is a trivariate Gaussian:
\begin{equation}
\begin{aligned}
\rm{P(V_{obs}, I_{obs}, Z_{obs}) = \frac{1}{\sigma_{V_{obs}}\sigma_{I_{obs}}\sigma_{Z_{obs}} (2\pi)^{3/2}} }\\
\times \exp(\frac{-(V_{obs} - V_{0})^{2}}{2\sigma_{V_{obs}}^{2}}) \\
\times  \exp(\frac{-(I_{obs} - I_{0})^{2}}{2\sigma_{I_{obs}}^{2}}) \\
\times \exp(\frac{-(Z_{obs} - Z_{0})^{2}}{2\sigma_{Z_{obs}}^{2}})
\end{aligned}
\end{equation}

where $V_{0}$ and $I_{0}$ are the model F606W and F814W magnitudes respectively and $Z_{0}$ is the model metallicity of the star. The $\sigma$ values are the observational errors on each measurement. We use a flat prior,  $ \rm P(\tau) = 1$ and consider age as the only free parameter. 

\begin{figure}
\centering
\includegraphics[width = .5\textwidth]{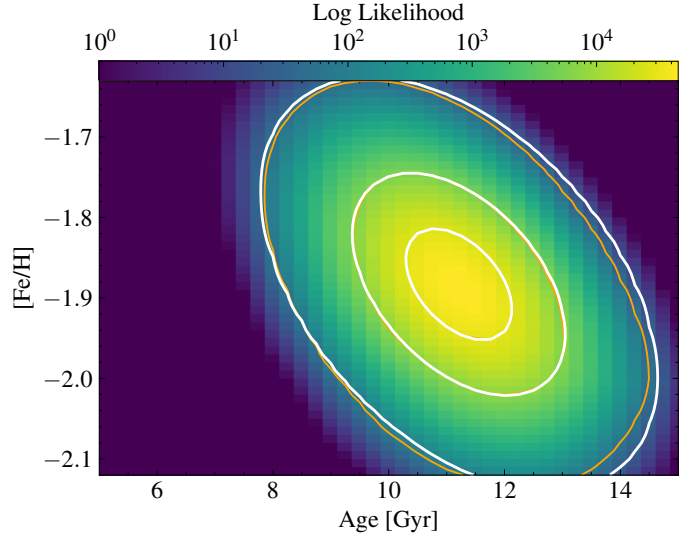}
\caption{\textbf{Posterior Probability:} This shows our sampling of the log-likelihood for one star in our selection. Overplotted are the PDF quantile levels (in orange) and Gaussian fit sigma contours (in white), showing excellent agreement at the 1 and 2 $\sigma$ levels.}
\label{fig:post_prob}
\end{figure}

We take a brute force grid sampling approach to evaluate the agreement of the data with isochrones.  First, we use the median error on [Fe/H] of our sample ($\rm [Fe/H]_{med} \sim 0.08$) and calculate the 3-sigma range ($\pm{0.24}$) and use this metallicity range as the window in which we sample the isochrone grid in [Fe/H] for each star in steps of 0.02 dex. With each [Fe/H] isochrone set we compare our star to the full range of model ages.  We calculate the log-likelihood for each interpolated point along a single set of model parameters, then take the sum of the log-likelihoods to find the posterior probability of the age and metallicity of that isochrone being a good fit for a given star.  We repeat this process for each age and for each sampling of [Fe/H] within the $3-\sigma$ error distribution to obtain the full posterior distribution function. We initially use the 50th percentile age of the Cumulative Distribution Function (CDF) to find the best-fit age.  Fig.~\ref{fig:post_prob} shows the posterior distribution function as a function of age and metallicity for one example star.   Some stars fall off our grid of models, and thus result in very low posterior probabilities. 171 stars have low maximum posteriors (set to $< 4000$) due to their best fit age falling well off the model grid sampled or the photometric and/or metallicity errors being large. For these stars we do not report an age constraint. 

We fit a 2D-Gaussian to each posterior probability so that we can capture the co-variance of the age and metallicity. The Gaussian fitting process has the advantage of allowing us to constrain ages that fall outside of the sampled grid. We find the best fit for the amplitude (A), central age ($x_0$), central metallicity ($y_0$), spread in age ($\sigma_x$), spread in metallicity ($\sigma_y$), and angle between the Gaussian terms ($\theta$) by minimizing the $\chi^2$ between the data and the model (g), given by:

\begin{equation}
\begin{split}
g & = A \cdot \exp{\Bigl[- \Bigl[ (\frac{\cos(\theta)^2}{2\sigma_x^2} +\frac{\sin(\theta)^2}{2\sigma_y^2})(x-x_0)^2}\\
& + 2 \cdot (\frac{\sin(2\theta)}{4\sigma_x^2} + \frac{\sin(2\theta)}{4\sigma_y^2})(x-x_0)(y-y_0)\\
& + \bigl(\frac{\sin(\theta)^2}{2\sigma_x^2} + \frac{\cos(\theta)^2}{2\sigma_y^2}\bigr)(y-y_0)^2\Bigr]\Bigr]
\end{split}
\end{equation}

Here x and y are the age and metallicity respectively. For initial guesses to this fit, we use the maximum log-likelihood and relevant quantile values of each of the age and metallicity normalized CDFs. For $\theta$ we set the initial guess to 0. Fig.~\ref{fig:post_prob} is an example of this process for one star in our sample. When comparing the orange contours (1, 2, and 3$\sigma$ levels of the Gaussian fit) to the white contours (1, 2, and 3$\sigma$ volume enclosed / quantile levels of the PDF), we see good agreement between the methods. While some stars exhibit a low probability tail to higher ages, we find that the Gaussian fitting provides a similar estimate of the 1$\sigma$ age uncertainty. 
Whenever the Gaussian 1$\sigma$ age uncertainty is lower than the $\sigma_g$  value calculated from the CDF (where $\rm{\sigma_g = 0.7413\times(75^{th} - 25^{th}}$ percentile)), we use the $\sigma_g$ value. Additionally, for any star where the Gaussian age uncertainty is below 0.25 Gyr, we set it to the age sampling, 0.25 Gyr, as these low errors result from grid coarseness. 
For our best-fit ages and errors we report the central age $x_0$ for each star and the spread in age $\sigma_x$ as the error.  

\begin{figure*}
\includegraphics[width =\textwidth]{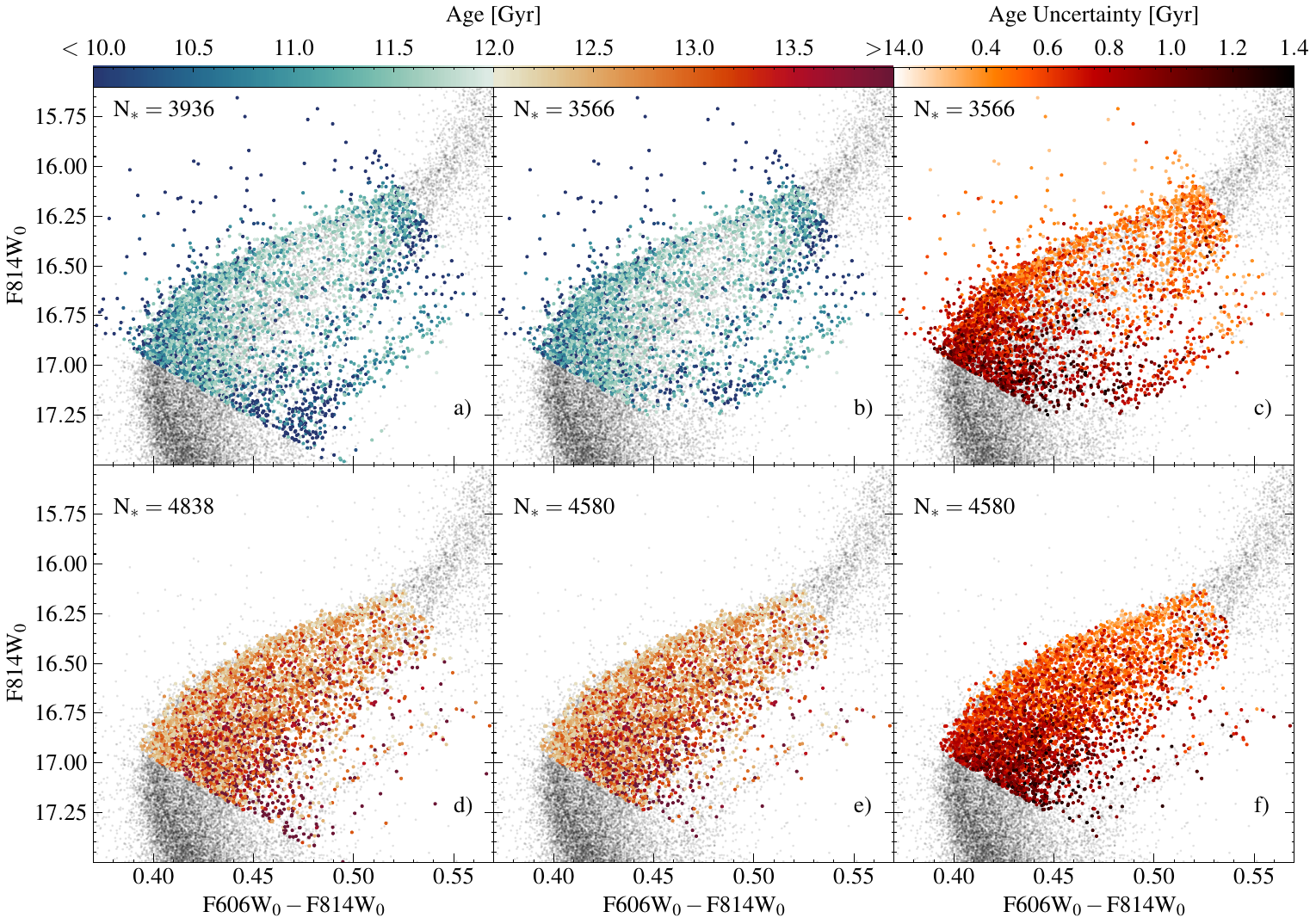} 
\caption{\textbf{SGB Ages and Age Uncertainties:} (\emph{all panels}) The full data set is shown by the grey points. (\emph{left panels}) The initial selection is shown colored by the age with the top panel showing all stars with ages less than 12 Gyrs and the bottom panel showing those with ages greater than 12 Gyrs. (\emph{center panels}) The reliable subsample after removing stars based on their age errors and systematics is shown colored by age. (\emph{right panels}) The reliable subsample is shown colored by the age uncertainty via Gaussian fitting.}
\label{fig:ages_cmd}
\end{figure*}

We initially calculate ages for the full SGB sample (9,129 stars) and show these measurements in the left panel of Figure~\ref{fig:ages_cmd}.  The errors are shown in the right panel.  To create a reliable sub-sample of SGB star ages we make additional cuts: (1) we remove stars with uncertainties of more than $> 1.4$ Gyr ($\sim$2$\times$ the median age uncertainty of 0.714 Gyr), (2) we remove stars with helium systematic age differences $> 1.4$ Gyr (full discussion in Section \ref{subsec:helium_systematics}) and
(3) we remove any stars for which we are not able to constrain all systematic ages ensuring that we can compare the final mean age and age spread results directly. The new selection is shown in the central panel of Fig.~\ref{fig:ages_cmd} and contains 8,146 stars. The selection primarily removes stars that reside near the MSTO and near the turnoff to the RGB due to larger errors in these regions. This selection also removes several of our youngest stars, especially amongst the metal-rich stars.  This is expected given the mismatch between the metal-rich isochrones at differing helium abundances as shown in Fig.~\ref{fig:sgb_region} (more discussion in Section \ref{subsec:helium_systematics}). The right-hand panel of Fig.~\ref{fig:ages_cmd} shows each star colored by its age uncertainty. 

\section{Stellar Age Results} \label{sec:stellar_ages}
To understand the star formation history of \omc\ we first find the ages of individual stars. Our choice to focus on the SGB region is discussed in detail in Section \ref{subsec:sample_selection} and the isochrone model comparison method is described in Section \ref{sec:methods}.

The ages range from 4.85 up to 16.29 Gyr with the median of the distribution being 12.16 Gyr. All ages fall within the age of the universe when considering their uncertainties. Comparison of Fig.~\ref{fig:sgb_region} and Fig.~\ref{fig:ages_cmd} shows that while the photometric sequences are distinct in metallicity, they are much less so in age, suggesting similar ages for many of the subpopulations in the cluster.  The exception is the most metal-rich stars, which have ages that are consistently younger than the median. We discuss the age-metallicity relation in the next subsection, followed by a discussion of the age spreads after accounting for errors in Section \ref{subsec:age_spread}. The catalog of primary and systematic age constraints is provided as a machine readable table (MRT) with this paper and a table describing its contents is provided in Table \ref{table:mrt_ages}.

\subsection{Age-Metallicity Relation} \label{subsec:age_metallicity_relation}
\begin{figure*}
\includegraphics[width = \textwidth]{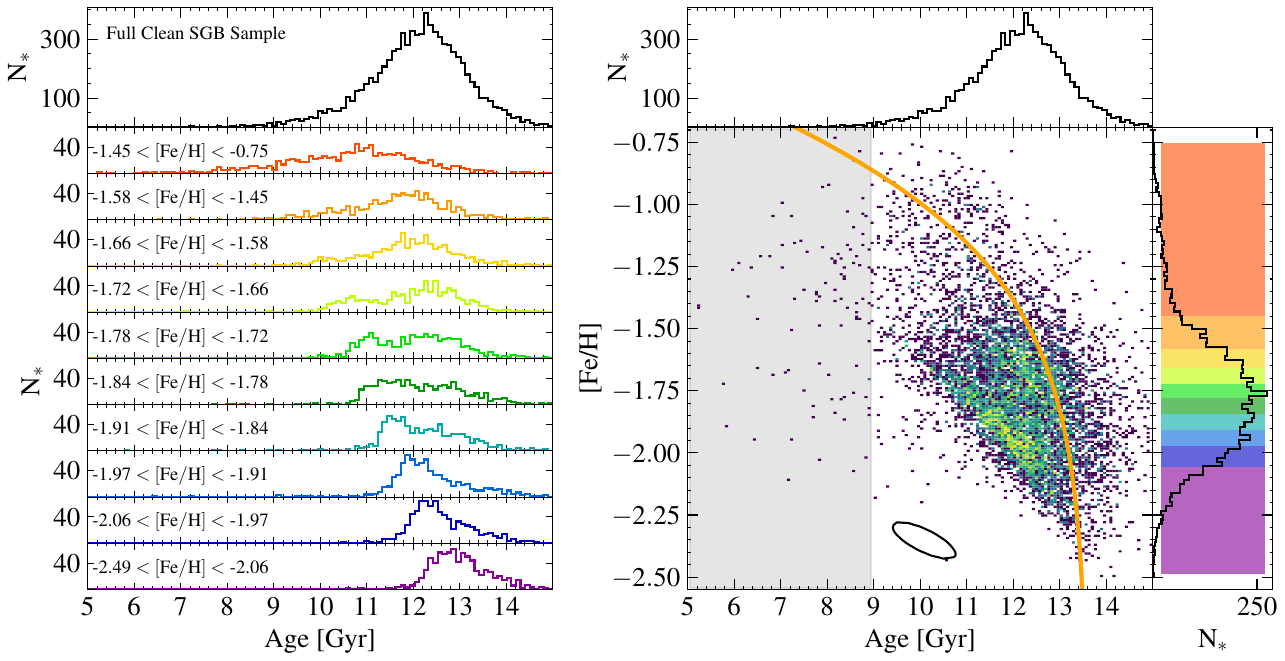}
\caption{\textbf{The Age-Metallicity Relation:} (\emph{left}) The age distribution of the SGB star sample in equal number ($ \rm N_* = 803$) metallicity bins. The colors correspond to the colormap of Figure 1. (\emph{right}) The 2D histogram of stars in the age-metallicity plane is given by the colormap, where purple is less dense and yellow is denser. The median uncertainty is given by the black contour in the lower center. The Gaia-Sausage /Enceladus (GS/E) SFH \citep{Limberg_2022} is shown by the orange line. On the right side we show the full [Fe/H] histogram with each equal number bin region denoted by colored patches. On the top of both panels we show the full age histogram for the clean SGB sample.}
\label{fig:age_met_relation}
\end{figure*}

We construct an age-metallicity relation for \omc\ by plotting the spectroscopic metallicity vs. the Gaussian-fit age for each star in our high quality sample in Fig.~\ref{fig:age_met_relation}. The left panel shows a histogram of the inferred ages in ten equal number bins in metallicity. We chose this binning scheme to ensure similar errors on all combined measurements. We have confirmed that implementing a metallicity binning scheme consistent with the Gaussian mixture model used in \cite{Nitschai_2024} does not alter our results. The right panel shows the age and metallicity of each individual star. The median uncertainty in age and metallicity from our Gaussian-fitting is given by the black contour in the lower left. The histogram in the left panel of Fig.~\ref{fig:age_met_relation} shows two clear trends: (1) the age of stars gets steadily younger with increasing metallicity with a mean age of 12.87 Gyr at the lowest metallicities and a mean age of 11.05 Gyr for the highest, and (2) the age spread of the stars grows with increasing metallicity. We show in the next subsection that the age spreads are significant at all metallicities.  A more careful examination of both panels in Fig.~\ref{fig:age_and_age_spread_systematics} shows an interesting additional trend; there is a clear bifurcation in the age-metallicity relation with a strong linear feature on the lower-left side of the diagram extending from older and metal-poor to younger and metal-rich.  This sequence is visibly separate from a more diffuse sequence that is more metal-rich at a given age; a clear gap in these sequences is visible at [Fe/H] $\sim$ -1.75, where the age histogram is clearly bimodal.  Separating these two sequences out over narrow ranges of metallicity shows that these stars occupy the left-most edge of the CMD at each metallicity.  The lower-left sequence contains $\sim$36\% of the stars at [Fe/H] of -1.75; this fraction increases towards lower metallicities (where the two sequences become less distinct), and decreases towards higher metallicities.
The number of stars falls dramatically off above [Fe/H]~$>$~$-$1.5, but these stars show distinctly younger ages, with the stars at [Fe/H]$\sim -1$ having $\sim$1~Gyr younger ages than those at lower metallicities.  
We discuss the interpretation of the two sequences in the age-metallicity relation and the younger ages at high metallicities in Section~\ref{subsec:interpretation_of_sfh}, and discuss additional stellar population information on the sequences in Section~\ref{subsec:future_work}.


At ages $<$9~Gyr there are a small number of stars (108; 1.3\% of all stars).  These cover a wide range of metallicities.  Looking back at Fig.~\ref{fig:ages_cmd}, we can see these stars typically fall above the main body of the SGB.  We suspect that at least some of these stars are contaminants: either evolving blue stragglers \citep[e.g.][]{Zhang_2021,Cerqui_2023} or binaries/chance superpositions of MSTO stars with fainter stars (or SGB stars with white dwarfs).  These contaminants would explain why these stars are found at a wide range of metallicities. We have therefore removed these stars from our age-spread analysis in the subsection. However, we note that relative to the total number of stars at each metallicity, there are many more of these stars amongst the more metal-rich populations. Specifically, while these young stars make up only 0.4\% of all stars below [Fe/H] of $-1.5$, they make up 7\% of the more metal-rich stars.  This suggests that at the metal-rich end, many of these stars may in fact be truly young members of \omc.  

To test for binaries we compare our young star sample to the \cite{Wragg_2023} binary catalog for \omc, which reports an overall binary fraction of $2.41 \pm{0.53}\%$. \cite{Bellini_2017c}, reports a similar binary fraction of $2.70 \pm{0.08}$\%, suggesting binaries are not a major contaminant in our sample. Upon comparison we find only 43 radial-velocity selected binaries lying within our SGB selection, of which 2 have ages below 9 Gyr (the youngest being 6.74 Gyr). This also supports the possibility that some of the young stars might have truly young ages.  The presence of these stars would be surprising, since most papers place the merger of the Gaia/Enceladus at $\gtrsim$8 Gyr \citep[e.g.][]{Belokurov_2018,Helmi_2018, Xiang_2022} ago and we would expect star formation to be minimal in the cluster after it was no longer at the center of a galaxy, though some residual gas and star formation may exist as is seen in for example in the common nuclear starbursts in early-type dwarf galaxies \citep{Koleva_2014}. An examination of the most metal-rich and youngest stars is planned for a future paper.

\subsection{Measuring the Age Spread} \label{subsec:age_spread}

In this section, we focus on constraining the age spread in the data. We do this by deconvolving the age distribution with the age errors assuming the age distribution follows a Gaussian distribution using a Gaussian likelihood function including heteroscedastic errors \citep{pryor_1993}. We then minimize the log-likelihood with the Markov-chain Monte Carlo routine {\tt emcee} \citep{Emcee_2013} to find the mean age and age spread of the full sample. Using our reliable star sample, the inferred age spread (after accounting for the errors and removing stars younger than 9 Gyrs) is 0.75 $\pm{0.01}$ Gyrs.  If this were due solely to underestimated errors, our errors would need to be scaled by a factor of more than 2.5 to remove the age spread.  However, we believe our age errors are robust as they are based on repeat photometric measurements (with a median of 62 and 14 measurements in F606W and F814W) and errors on the metallicity that have been scaled based on repeat measurements \citep{Nitschai_2023}.  

\begin{figure*}
\includegraphics[width = \textwidth]{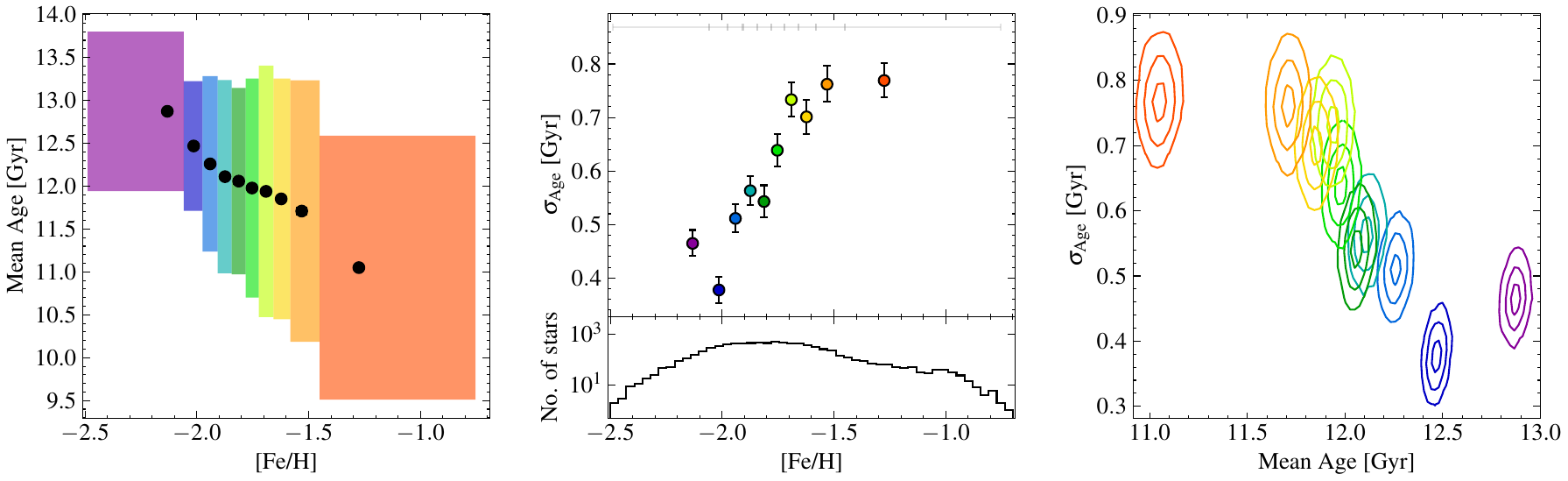}
\caption{(\emph{left}) \textbf{Mean Age and Age Spread:} Our mean age and deconvolved age spread constraints are shown in black circles and blue boxes respectively. The black points are located at the median [Fe/H] of the bin. (\emph{center}) \textbf{Age spread vs. [Fe/H]:} The lower panel shows the metallicity histogram for our sample. The upper panel shows the age spread for each metallicity bin. The gray lines across the top delineate the extent of our equal number metallicity bins. (\emph{right}) \textbf{Age spread vs. Mean Age:} The contours are color-coordinated with the left panel to indicate metallicity. We show the 2D 1, 2, and 3 $\sigma$ confidence levels from the MCMC fitting for these values. The age spread increases with age.}
\label{fig:age_vs_age_spread_contour}
\end{figure*}

We then use our reliable SGB star sample and create ten [Fe/H] bins with equal numbers of stars after rejecting the youngest stars, which we argue above may be contaminants.  Each metallicity bin contains $\sim$800 stars.  In the left panel of Fig.~\ref{fig:age_vs_age_spread_contour} we show the mean age as a function of metallicity by the black points. The age spread at each metallicity is illustrated by the blue boxes, centered at the mean age. In the central panel, we show the age spread vs. the mean metallicity for each sample. Interestingly, even the more metal-poor stars have an age spread of around 0.40 Gyr, implying that at least two distinct metal-poor populations must have formed at different times, and then coalesced as might be expected from the inspiral of at least two clusters. This finding agrees well with the extended (and complex) age distribution we see for even the low metallicity stars in the left panel of Fig.~\ref{fig:age_met_relation}.  The age spread increases significantly with metallicity and the metal-rich stars have by far the highest age spread. The overall trend of increasing age spread with metallicity flattens around [Fe/H]=$-1.6$, with more metal-rich bins having nearly equal age spreads. We see a clear decrease in age spread at [Fe/H] $= -2.0$ relative to bins at both higher and lower metallicity. This could be due to a simpler stellar population at this metallicity; either there could be a large single stellar population as might be expected from a globular cluster infall event, or it could be due to the bimodality in the age-metallicity diagram disappearing at these lower metallicities (although this would not explain the increase in the lowest metallicity bin).  

The right panel of Fig.~\ref{fig:age_vs_age_spread_contour} shows the age spread vs. age confidence contours for the same metallicity bins, sharing the color scheme with the central panel. Here we can see that although the spread levels out at metallicities about $-1.6$, the mean age continues to get younger in these bins.  The monotonic decrease of mean age is also seen in the right panel of Fig.~\ref{fig:age_met_relation}. 
We note that some contribution to the intrinsic age spread may be due to systematic errors that contribute to the uncertainties in our age estimates. Specifically, the variations in helium or $\alpha$ abundance within a given metallicity bin will contribute to the age spread we infer, but as we show in  Section \ref{sec:systematics}, this appears to be a small effect, much smaller than the spreads we see in age. 
 
\section{Robustness of Results: Analysis of the Systematic Uncertainties} \label{sec:systematics}
All isochrone models are built upon the current understanding of the complex process of stellar evolution and include an array of model parameters known to affect the CMD position of a star throughout its lifetime. This means we have certain model assumptions that affect our stellar age determinations.  Of particular note in \omc\ are the presence of the C+N+O variation with metallicity, the $\alpha$ abundances, the spread in helium, and uncertainties in the distance to the cluster.

We constrain the contribution to the errors on our age estimates due to various of these sources of uncertainty, including those due to model assumptions. We divide these into two categories; those where we have made model choices that may be different from previous works, and intrinsic uncertainties that contribute to our observations, specifically our inability to constrain the helium abundances of individual stars despite known variation and the distance uncertainties.

\subsection{Systematics Due to Model Choices}

\begin{figure*}
\includegraphics[width = \textwidth]{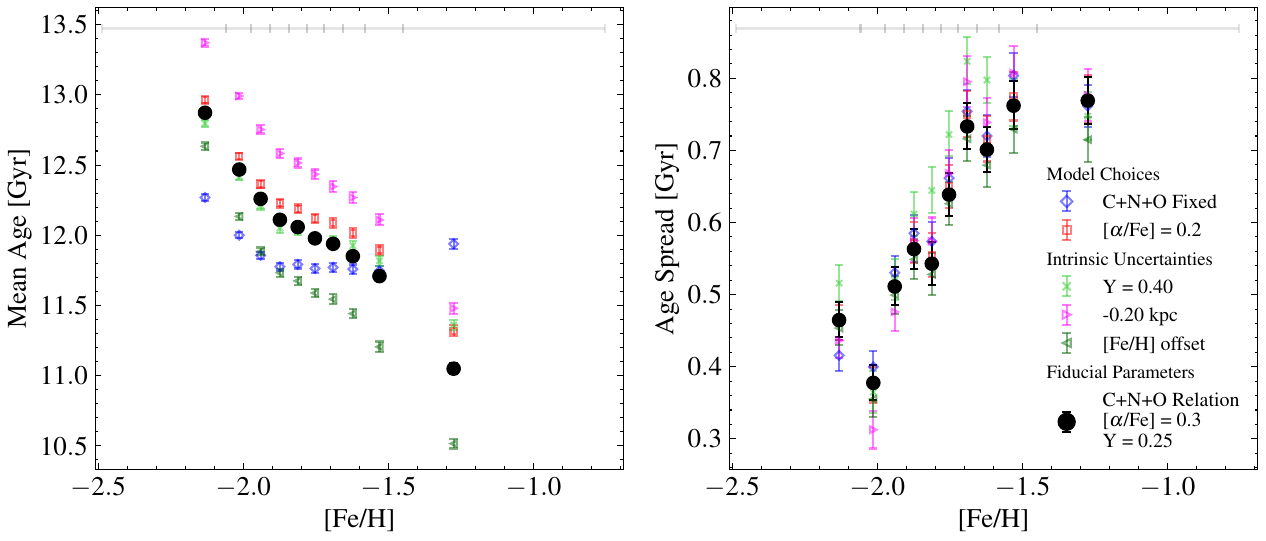}
\caption{(\emph{left}) \textbf{Mean Age vs. [Fe/H] Systematics:} Our mean age (\em{left}) and age spread (\em{right}) as a function of metallicity with our fiducial results (black points) as well as systematic tests (colored points; marker labels shared with both panels).  In both panels errorbars show 1$\sigma$ uncertainty and gray bars along the top denote the [Fe/H] equal number bin edges. Generally, our systematics tests slightly affect the mean ages but not the trend with [Fe/H] except for the C+N+O abundance, which significantly impacts this measurement. The age spread is largely independent of any systematic effects.
}
\label{fig:age_and_age_spread_systematics}
\end{figure*}

\subsubsection{C+N+O variation with [Fe/H]} \label{subsec:cno_fe_systematics}
The largest model improvement undertaken for this project was generating isochrones with a C+N+O vs. [Fe/H] relation specifically for \omc. It is therefore very important to understand the effects this has on our age determinations when compared to C+N+O-fixed models. We run age determinations with the C+N+O fixed models, keeping all other model parameters the same as our fiducial set. These systematic ages are shown in Fig.~\ref{fig:age_and_age_spread_systematics}.  If we don't incorporate the C+N+O variations we find a much smaller difference in ages as a function of [Fe/H], with a nearly constant age-metallicity relation from [Fe/H] of $-$2 to $-$1.5, and with the highest metallicity stars actually being older than those at lower metallicities.  

\begin{figure}
\includegraphics[width = .45\textwidth]{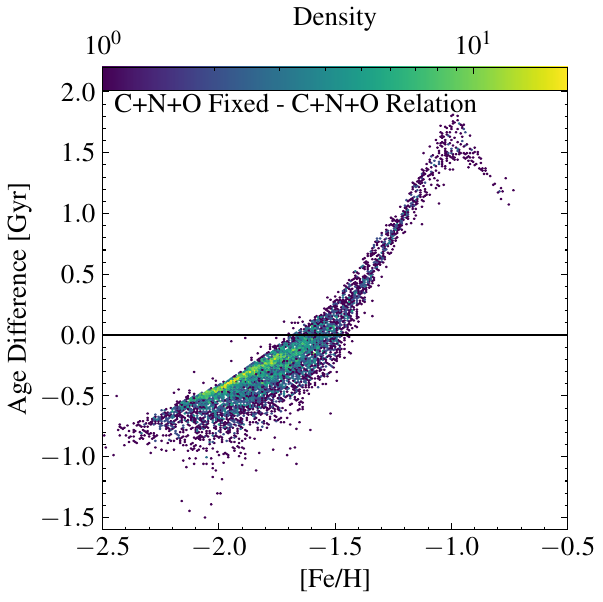}
\caption{\textbf{C+N+O Systematics:} The consideration of the relation between C+N+O enhancement and [Fe/H] makes a noted difference in age determination, though still largely within two times the average age error. The turnover at [Fe/H] = -1.0 is due to the C+N+O relation being held at a constant value above this metallicity.}
\label{fig:cno_systematics}
\end{figure}

We show a direct comparison of the ages with our new models with the C+N+O relation relative to the C+N+O fixed models in Fig.~\ref{fig:cno_systematics}.  
As expected, the difference in age determination is a strong function of metallicity with the C+N+O fixed ages being younger at low metallicity and significantly ($\sim$1.8 Gyr) older at the highest metallicities. The similarity at [Fe/H]$\sim -1.7$ is expected as the C+N+O value in our relation is equivalent to the $\rm[\alpha/Fe]=0.3$ in the C+N+O fixed isochrones.  
The high metallicity inflection is due to our C+N+O enhancement being held at a constant value at metallicities higher than [Fe/H] = $-$1.0 due to a lack of data in the \citet{Marino_2012} paper at the highest metallicities. We note that the \citet{Marino_2012} paper found minimal scatter of the C+N+O abundance at a fixed metallicity, thus we don't think that this relation provides a significant additional source of scatter in our derived ages. \citet{Marino_2012} noted a potential small $\sim$0.1 dex  difference in the C+N+O between first and second generation stars at a fixed metallicity (which is not included in our fitted relation).  This difference is too small to account for the large differences in age we see in the two sequences in Fig.~\ref{fig:age_met_relation}, and would work in the opposite direction to the age difference we see between the populations (assuming the younger population is second generation; see Section~\ref{subsec:future_work}).  

Despite the large difference in the inferred mean ages, the age spreads shown in the right panel of Fig.~\ref{fig:age_and_age_spread_systematics} are in excellent agreement between the two models with different C+N+O prescriptions at all metallicities.

\subsubsection{$\alpha$ abundance variations} \label{subsec:alpha_systematics}
To test our systematic uncertainties due to our choice of alpha abundance we calculate ages with $\rm[\alpha/Fe] = 0.20$ for the full range of metallicities and plot these as red square markers in Fig.~\ref{fig:age_and_age_spread_systematics}. This difference with our reference value of $\rm[\alpha/Fe] = 0.30$ is similar to the intrinsic scatter inferred in the [Si/Fe] abundances in \omc\ from \citet{Johnson_2010}.
We find good agreement in the age determinations compared to our fiducial model, with differences increasing slightly (from 0.09 to 0.27 Gyr) across the metallicity range. 

The $\rm[\alpha/Fe] = 0.20$ age spread constraints are well within $1\sigma$ from the fiducial models at all metallicities, showing relatively uniform increase (over fiducial) in age spread with the average being $\sim 0.02$ Gyr. These comparisons together definitively show that our alpha abundance model choice is not a major contributor to our systematic uncertainties, nor do we expect star-to-star variations in [$\alpha$/Fe] to add significantly to the inferred age spreads. 

\subsubsection{Using F625W instead of F606W} \label{subsec:f625w}
Our \textit{HST} photometric catalog provides F625W magnitudes over a much wider spatial area than the available F606W magnitudes; these measurements cover stars out to 9.0 pc in some regions (with the average radius being 4.2 pc). The total number of SGB stars in F625W is 18,352. However, the quality of the individual photometric measurements is of much lower quality. Due to saturation in the longer F625W exposures, 86\% of SGB stars have photometry based on just a single exposure.  This means that the photometric errors are fairly high (median of 0.026 mag) and that the possibility of catastrophic outliers is much higher (due to detector issues or cosmic rays).  We therefore choose to use data from the F606W filter, where we have a median of 62 separate photometric measurements and median errors of 0.006 mag. 

While the overall number of stars observed is lower in the F606W filter, this band's increased photometric precision means age constraint errors are significantly lower (0.7 Gyr as compared to 2.0 Gyr). However, the F606W measurements are focused only on the inner 5 pc, with a median radius of 2.5 pc. 

To use F625W as a systematic check we use both the same selection as for the F606W analysis and the full F625W sample ($N_* = 16k$), and rerun our age and age spread constraints (shown in Fig.~\ref{fig:f625w_systematics} by purple and orange star markers respectively.  The ages constrained by both F625W are all systematically younger than the fiducial ages, with the F606W sample showing a larger mean age decrease. This leads us to conclude that filter choice does have a significant impact on our absolute age determinations, likely due to zeropoint offsets between the two photometric bands. Though, because the systematic offset is relatively uniform, it has no impact on relative ages, as demonstrated by the agreements seen among the age spread constraints in the right panel. This also allows us to conclude that our errors are appropriately estimated. It is also worth nothing that because the full F625W sample extends to a much larger radius, we can conclude there is no strong gradient in age with radius at least within the half-light radius of \omc.

\subsection{Systematics Due to Intrinsic Uncertainties}

\subsubsection{Helium}\label{subsec:helium_systematics}
\begin{figure}
\includegraphics[width = .45\textwidth]{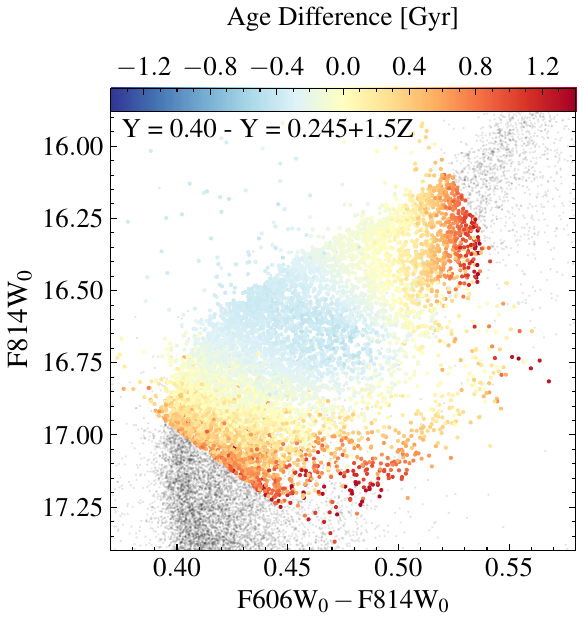}
\caption{\textbf{Helium Abundance Systematics:} Each point in our final selection is colored by the difference in age determination when considering helium enhanced isochrones vs. solar. Our selection is limited to stars with a helium age difference of less than two times the average age error which only removes 216 stars.}
\label{fig:helium_systematics}
\end{figure}

Our initial stellar ages are calculated using isochrones with Y$= 0.245+1.5Z$.  However, the stars in \omc\ are known to have a range of helium abundances with some as high as Y$=$0.4 \citep{Reddy_2020} ($\Delta Y=0.15$). We calculate ages with helium enhanced isochrones and show the differences in the resulting age determination on the CMD when compared with our primordial helium models (keeping all other parameters the same) in Fig.~\ref{fig:helium_systematics}. The differences closely follow the offsets between the model isochrones that can be seen at two metallicities in the central and right panels of Fig.~\ref{fig:sgb_region}. Model differences increase along the MSTO as well as approaching the RGB region where the helium enhanced isochrones differ most from the lower helium abundance isochrones. 

We also plot the helium-enhanced mean age and age spreads in Figure~\ref{fig:age_and_age_spread_systematics} with limegreen ``x'' markers. The differences are small at lower metallicities, with the ages being slightly younger than the fiducial ages at the lowest metallicites, and then slightly older at higher metallicities. The only large difference is seen in the highest metallicity stars where the helium abundance is known to be significantly enhanced \citep[][Clontz et al. {\em in prep}]{Tailo_2016}. It is worth noting that helium is the model assumption that has the least effect on our age determinations. Even if we consider that the Y$=$0.4 helium abundance may provide more accurate ages than the fiducial ages, we would still find a significant, monotonic age-metallicity relation extending to the highest metallicities.

When looking at the age spread we do not see deviations greater than $2 \sigma$ except for the highest metallicity bin and no relation with metallicity. While helium does have a small effect on age determination (and would therefore affect the overall age spread of the cluster), when considering age spread in narrow metallicity ranges, it has no significant impact on our measurement.  The age spread could be enhanced somewhat from a spread in helium at a given metallicity, however, this effect is limited since the mean ages inferred are very similar across the full range of expected helium abundances.  

\subsubsection{Distance} \label{subsec:dist_systematics}
We also rederive the SGB stellar ages to understand the systematic impact of the distance uncertainty.  We recalculate the values at a smaller distance ($\rm \Delta D = 0.20 kpc$) of 5.23 kpc (giving ($\rm{m}-\rm M_0$) =  13.590 and $\rm \Delta DM = -0.082$)).  This is a generous uncertainty on the distance, larger than the quoted uncertainty in \cite{Baumgardt_2021} of $\pm{0.05}$ kpc.  Instead, our chosen values reflect the overall spread in literature values, including the independent Gaia distance estimate from \citet{Soltis_2021} of 5.24$\pm$0.11~kpc. Because age and metallicity are degenerate on the CMD, it is expected that by decreasing the distance we would increase the age and that this would not vary with metallicity. This can be seen in Fig.~\ref{fig:age_and_age_spread_systematics} when comparing the distance systematic ages (shown by pink right-facing triangles) to our fiducial model ages. We also do not see notable difference in the age spreads due to distance uncertainties which is consistent with all ages simply being offset equally. While distance does play a significant role in age determinations, it is not a major contributor to uncertainty in our age spread measurements or the relative ages between the different metallicities, only on the absolute age estimates. We note that the decreased distance makes the oldest stars uncomfortably old relative to the age of the Universe.

\subsubsection{Individual C, N, O, Mg, Al, and Si variations}
Even though the sum C+N+O remains constant for a metallicity subpopulation, the individual abundances of C, N, O, Mg, Al, and Si change star-by-star \citep{Bastian_2018,Milone_2022}. Therefore, to investigate the impact of these variations on the age determination, we follow the same approach as described by \cite{Dotter_2015}, \cite{Milone_2018}, and \cite{Lagioia_2019} to derive the expected magnitude difference by changing the stellar chemical abundances.
We built a pair of synthetic spectra for each metallicity bin using the codes ATLAS12 and SYTNHE \citep{Kurucz_1970,Kurucz_1993,Sbordone_2004}. To select the effective temperature and surface gravity to construct the synthetic spectra, we used the equivalent evolutionary phase (EEP) around 160 since it represents the SGB mean locus very well. One spectrum has a solar composition with the updated values from \cite{Asplund_2021}, with, however, $\alpha$-enrichment being [Mg/Fe], [O/Fe], and [Si/Fe] equal to +0.40. The other synthetic spectrum is built to simulate the maximum chemical abundance variation between the first (1G) and second (2G) generations: [C/Fe]$=-0.5$, [N/Fe]$=+1.5$, [O/Fe]$=-0.10$, [Mg/Fe]$=-0.10$, [Al/Fe]$=+1.5$, [Si/Fe]$=-0.10$. As noticed by \cite{Cassisi_2013} and \cite{Milone_2018}, the variation in Mg, Al, and Si contributes negligible amounts to the magnitude differences, however, we included them in our test for completeness. We integrate the spectra over the band-passes of all \textit{HST} filters employed in this work to create synthetic magnitudes, and then these are compared between the 1G and 2G spectra. 
For the two colors used to derive SGB ages in this work (F606W-F814W and F625W-F814W), the predicted absolute differences are smaller than $0.005$ mag. This difference is smaller than the photometric color errors on our SGB stars and translates to an age difference of $< 0.10$~Gyr, much smaller than our median age error. Therefore, the star-by-star light element abundance variations are not a substantial source of uncertainties in our age determination method. 

\subsubsection{Metallicity Offset}\label{subsec:metallicity_offset_systematic}

The metallicity distribution of the SGB stars shown in Fig.~\ref{fig:age_met_relation} is not fully consistent with the metallicity distribution of the RGB, as shown in Figure 3 of \cite{Nitschai_2024}. The median SGB metallicity is offset (compared to the median RGB metallicity) in the metal-poor direction by 0.064 dex. This is less than the median $\rm{[Fe/H]}$ uncertainty for the SGB stars (0.08 dex) but is significant enough to consider its systematic impact on our age estimates. This offset is likely due to dredge up in RGB stars increasing the inferred metallicity of these stars in comparison to SGB stars which have not yet begun dredging up heavy elements. We note that our metallicity estimates ($\rm{[Fe/H]}$ inferred from a spectroscopically measured $\rm{[M/H]}$) include the impact of many heavy elements 
\citep[see][for details]{Nitschai_2023}.  So while we think our SGB $\rm{[Fe/H]}$ estimates are reliable, we quantify the effect this $\rm{[Fe/H]}$ offset could contribute to our age uncertainties.  To do this, we take the median $\rm{[Fe/H]}$ offset between the RGB and SGB and add it to the $\rm{[Fe/H]}$ of each star before calculating its age. We then isolate the same set of stars used in our clean sample for comparison. We find the increased $\rm{[Fe/H]}$  decreases the ages of stars by a median value of 0.40 Gyrs, which is less than our median age error of 0.67. This offset does increase with metallicity, from 0.2 to 0.8 Gyr. The offset in ages does not affect our age spread constraints. The results are shown as dark green left-facing arrows in Figure \ref{fig:age_and_age_spread_systematics}

\section{Discussion}\label{sec:discussion}

We perform our MCMC routine to deconvolve the age distribution with the age uncertainty to obtain the mean age and age spread for our reliable SGB sample (still excluding stars with Ages $< 9$ Gyr. We find a mean age of $\rm 12.08 \pm{0.01}$\,Gyr and an overall age spread of $\sigma_{\rm age} = 0.75 \pm{0.01}$\,Gyr.

With the analysis presented here, we confirm that the large spread in metallicity in \omc\ is accompanied by a large spread in ages. We also show through this work that a significant age spread persists at fixed metallicity, only explainable through complex/multiple enrichment pathways, suggesting that stars formed in multiple locations before becoming part of the same cluster.

We discussed in detail our systematics due to model choices and intrinsic uncertainties in Section \ref{sec:systematics}. In addition to these, we also investigated potential contaminants in our sample, including binaries and chance superpositions. 

\subsection{Comparison to Literature}\label{subsec:comparison_to_literature}

Similar to \cite{Hilker_2004}, we find that age and age spread increases linearly with metallicity. We also find that stars with [Fe/H] $> -1.3$ are younger and that enrichment ends around [Fe/H]$ = -1.0$. They suggest an overall age range of 3 Gyr. We also see an age range of $\sim 3$ Gyr across the bulk of our population.

Our results qualitatively agree with \cite{Villanova_2014} who also reported a two-stream age-metallicity relation in their SGB ages study for \omc\ with the more metal-poor sequence actually being younger than the metal-rich at a fixed metallicity. We note that they do not derive absolute ages and our metallicities are 0.5 dex more metal-poor than their measurements. Regardless, they also note the peculiar and seemingly inexplicable nature of these two branches seen in the AMR and confirm that changing C+N+O abundance does not erase this separation. 

The age difference between the mean age of our oldest and youngest sets of stars is $\sim -1.9$ Gyr which is close to the age difference found by \cite{Joo_2013} of $\sim -1.7$ Gyr. They derived the ages for five separate full CMD sequences using literature estimates for the metallicity and abundances.  They also find their populations are coeval when they fix C+N+O to a constant value, in agreement with the much reduced (but still significant) age spread that we find (blue points in Fig.~\ref{fig:age_and_age_spread_systematics}).

When we perform the deconvolution for the full sample we get an age spread of $\sim 0.75 \pm{0.01}$ Gyr which is at odds with the coeveal ($\sigma_{age} < 0.5$ Gyr suggested by \cite{Tailo_2016}. They achieve a low age spread by enhancing helium (up to Y=0.37) and C+N+O abundance (up to [(C+N+O)/Fe] = 0.7) in their models for the most metal-rich populations. Interestingly, we see the opposite effect in our results (see Fig.~\ref{fig:age_and_age_spread_systematics}), where an increase in C+N+O (with helium fixed) for only the metal rich stars would increase the overall age spread. They are clear about adopting models that support their goal of demonstrating the feasibility of a coeval model and do not assert that their choices are the only ones that reproduce the CMD of \omc.

\subsection{An Initial Look at Subpopulation Ages}\label{subsec:future_work}
\begin{figure*}
\includegraphics[width = \textwidth]{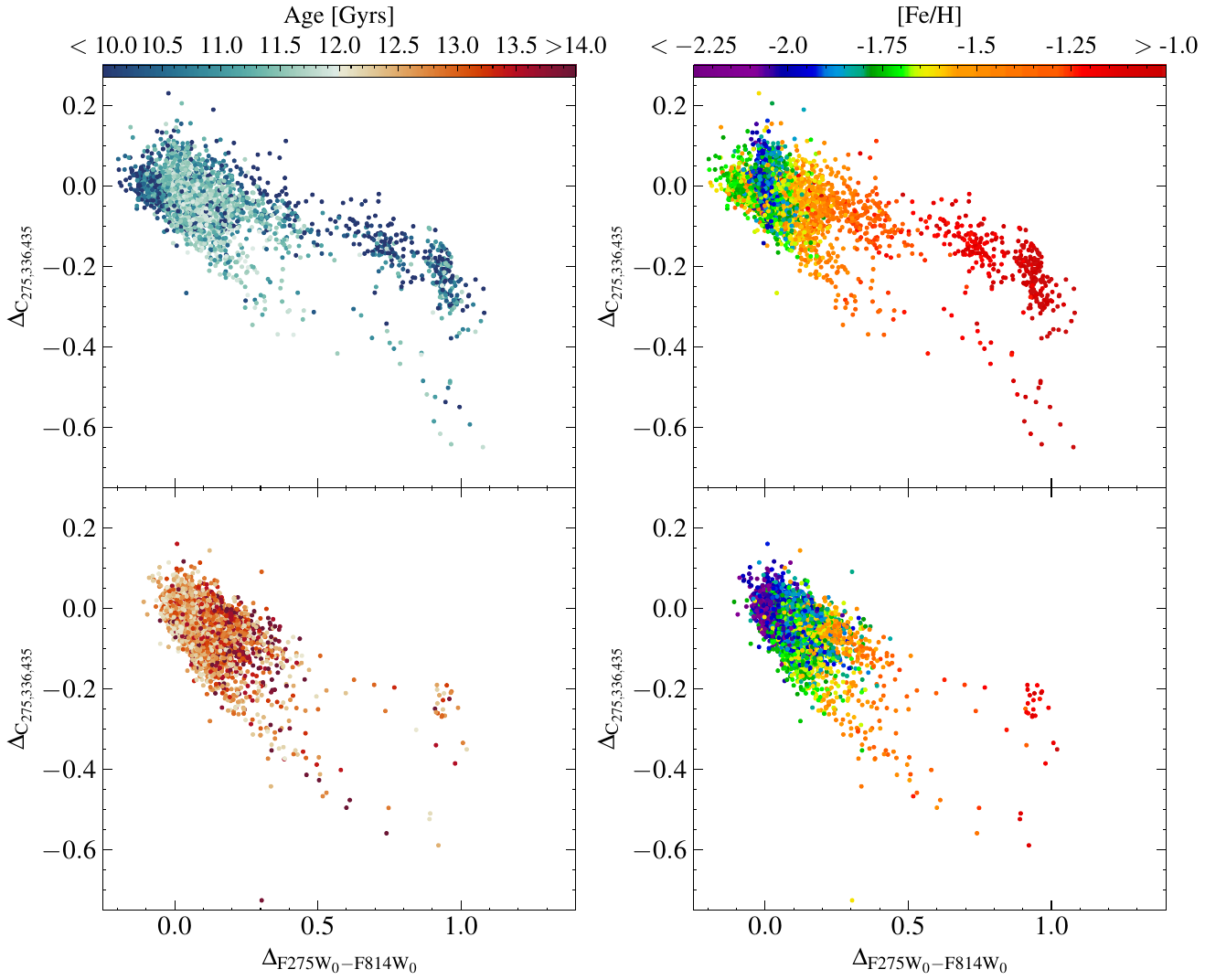}
\caption{\textbf{SGB Chromosome Maps:} (\emph{left panels}) Each star is colored by its age showing the distribution of old and young stars within the color-color space. (\emph{right panels}) SGB Chromosome Map: Each star is colored by its metallicity. We see several distinct features (see text for more details) as well as discrete subpopulations which will be the focus of future work.}
\label{fig:sgb_chromosome_map}
\end{figure*}

We also use our ages to examine the ages of subpopulations. We create an initial ``chromosome map" \citep[following methods outlined in][]{Milone_2017a} using the \textit{HST} multi-band photometry from \citet{haeberle_2024a} and coloring it using the SGB ages derived here (left panel of Fig.~\ref{fig:sgb_chromosome_map}). We also present the SGB chromosome map colored by metallicity (right panel of Fig.~\ref{fig:sgb_chromosome_map}), for comparison. Two features are immediately apparent.  First, there is a population of young stars on the left edge of the diagram; these stars are intermediate metallicity stars and likely form a distinct subpopulation. This group corresponds to the upper-left region of the lower stream in our age-metallicity relation which has a notable age offset to younger age compared the upper stream at this metallicity. Second, the abundance enhanced/second generation stars at high metallicities ($\rm \Delta_{F275W_0 - F814W_0} = 1$ and $\rm \Delta C_{275, 336, 435} = -0.2$) are significantly younger (greater than the age error, on average) than the first generation branch at the same metallicity ($\rm \Delta C_{275, 336, 435} = -0.5$).  This difference may be enhanced by the likely difference in helium amongst these stars, but the observed difference is much larger than the $\sim$0.3 Gyr difference seen with varying the helium abundance in Fig.~\ref{fig:age_and_age_spread_systematics}.  

We can also draw comparisons between the metallicity colored SGB chromosome map and the metallicity decomposition RGB chromosome map presented in Figure 18 of \cite{Nitschai_2024}. In both cases we see a lower and upper stream, typically referred to as the 1G (first-generation) and 2G (second-generation) sequences respectively. The RGB map shows a more distinct third stream at the lower metallicities, lying between 1G and 2G while our map shows a squeezing of these sequences on the SGB. Another difference is the metallicity gradient that is notable in the RGB chromosome map is less apparent on the SGB due to light element abundances more greatly affecting the slope of stellar population sequences here causing them to no longer lie parallel to one another, meaning they overlap in the chromosome map space. Nonetheless, we can examine the relative positions of the two streams we have identified in the age-metallicity relation.  In general, the younger, lower-left sequence in the age-metallicity relation is found to the upper left of stars with similar metallicity on the upper-right sequence in the chromosome map.  This displacement is typically associated with enriched 2G populations.  The displacement along the x-axis may due to just the age gap in the stars,  however, the accompanying offset on y-axis suggests that there may also be light-element abundance differences and that the lower-left sequence is a 2G sequence. Resolving  these possibilities would be facilitated by connecting the SGB populations here to populations along the RGB and MS \citep{Milone_2017b,Bellini_2017c}.  This is a challenging effort and will be the topic of an upcoming paper (Clontz et al., {\em in prep}).

\subsection{Interpretation of the Complex Age-Metallicity Relationship in \omc}\label{subsec:interpretation_of_sfh}

Our SGB ages reveal the time between the formation of populations with varying metallicities (and abundances).  These age differences can yield insight into the formation and chemical enrichment history of \omc.
The clear trends in the age-metallicity relationship shown in Fig.~\ref{fig:age_met_relation} as well as the significant spreads in the age distribution at each metallicity shown in Fig.~\ref{fig:age_vs_age_spread_contour} show how complex \omc's assembly history is. The mean age varies by more than a Gyr from the youngest to the oldest ages and this variation in the mean age is highly significant with a mean age of 12.87 Gyr at the lowest metallicities and a mean age of 11.05 Gyr for the highest, while the age spread reaches $\sim$0.8 Gyr for the most metal-rich populations and may be even larger given the number of very young stars amongst this population have been removed for this analysis.

Of particular interest for understanding \omc's formation are the two clear sequences seen in the age-metallicity diagram (Fig.~\ref{fig:age_met_relation}).  The presence of multiple metallicity populations at a single age is a clear indication of multiple sites of star formation.  The narrow age-metallicity track of the lower-left population meets the expectations of a long period of uninterrupted self-enrichment \citep[e.g.][]{Xiang_2022}. Thus this might indicate that this population formed {\em in situ} within the nuclear star cluster.  At the same time, the upper-right population has an age-metallicity relationship that is similar to the one inferred for Gaia-Enceladus by \citet{Limberg_2022} using the globular cluster population associated with that proto-galaxy.  It would therefore make sense for this to be an accreted population of inspiralling globular clusters into the Gaia-Enceladus nuclear star cluster, which later was stripped to become \omc.  Both {\em in situ} and globular cluster inspiral are expected in NSCs \citep{Neumayer_2020}.  Furthermore, NSCs in galaxies with stellar mass around that inferred for Gaia-Enceladus \citep[$\sim$10$^9$~M$_{\odot}$][]{Limberg_2022} are particularly likely to have populations from both globular cluster inspiral and {\em in situ} formation \citep{Fahrion_2021}. 

However, this interpretation comes with at least two significant challenges.  First, the lower-left, potentially {\em in-situ} component is more metal-poor at a fixed age than the upper-right, potentially accreted component.  This is exactly the opposite of expectations for star formation within a galaxy, where the central regions would be expected have the higher metallicity at any given age, not lower. One possible explanation is that these stars formed from accreted pristine gas from a merger event. Second, there is evidence of continuing enrichment in \omc.  Most notably, the fairly tight, monotonic relation in C+N+O seen by \citet{Marino_2012} would not be expected if roughly half the stars in \omc\ were formed in an environment that wasn't undergoing ongoing self-enrichment.  However, we note that the \citet{Marino_2012} data comes from stars at a much wider range of radii than we study here, and thus known gradients in the stellar populations \citep[e.g.][]{Bellini_2009, Calamida_2020} could be different between our SGB sample and the abundances studies in \citet{Marino_2012}. Based on the chromosome map (Figure~\ref{fig:sgb_chromosome_map}) the lower-left sequence appears to be an enhanced second generation population.  This would suggest a large age gap between the first and second generation stars at intermediate metallicities that is in clear tension with the similar C+N+O seen for all stars at a given metallicity by \citet{Marino_2012}.  Directly testing the abundances of SGB stars on these two sequences in the age-metallicity plane would provide a path forward for understanding their origin.

\section{Conclusions} \label{sec:conclusions}
We present age determinations for over 8100 SGB stars in \omc\ by combining high precision \textit{HST} photometry and MUSE spectroscopic metallicities.  We use these ages to examine the age-metallicity relation; the most comprehensive age-metallicity relation created for \omc. We find:

\begin{itemize}
    \item  The bulk of the stars have ages between 13 and 10 Gyr, with a mean age of $ \rm 12.08 \pm{0.01}$ Gyrs and an overall intrinsic age spread of $0.75 \pm{0.01}$ Gyr.

    \item The age-metallicity plane shows two distinct populations at low metallicity; a lower-left, younger population with a clear age-metallicity relation, and an upper-right older population with a less distinct age-metallicity relation (Fig.~\ref{fig:age_met_relation}).  At fixed age the lower-left is more metal-poor than the upper-right population. The upper-right population shows better agreement with the age-metallicity relation of other globular clusters associated with Gaia-Enceladus derived in \citet{Limberg_2022}.  
    \item The mean age decreases monotonically with increasing metallicity. The age spread also increases with age, but levels out at [Fe/H] of $\sim -1.6$ and age of $\sim 11.5$ Gyr (Fig.~\ref{fig:age_vs_age_spread_contour}).  

    \item Our age-metallicity results are robust to any likely systematic errors due to helium or $\alpha$ abundance variations, while distance uncertainties impact primarily the absolute age, but not the other age-metallicity trends. Not accounting for the observed C+N+O vs. [Fe/H] relation results in a much smaller spread in mean ages (Fig.~\ref{fig:age_and_age_spread_systematics}).  
        
    \item The chromosome map shows clear features that indicate populations with distinct ages (Fig.~\ref{fig:sgb_chromosome_map}).  
    
\end{itemize}

Our age constraints for stars on the SGB builds on the method used for M54 by \cite{Alfaro-Cuello_2019}, though the number of age dated stars in our sample is larger than in any other cluster. In the future, we will combine the ages with subpopulations from photometric chromosome map (Clontz et al. in prep) and individual stellar chemical abundances from MUSE spectra (Wang et al. in prep) to comprehensively study the chemical evolution and assembly history of \omc.

The results presented here show that \omc\ gives us a rich view into how NSCs are assembled and evolve. Its origin, assembly history, and star formation history continue to demand careful consideration to disentangle. 
As more details surrounding the formation of \omc\ become clear, the data point toward a complex star formation history that is consistent with the multiple generations and pathways of formation seen in NSCs, providing additional evidence that \omc\ was once the nucleus of a building block of the Milky Way.

\section{Software and third party data repository citations} \label{sec:software}

\begin{acknowledgments}

We would like to thank the referee for their helpful comments which greatly improved this work.
CC acknowledges the contributions to this work via the high performance computing resources at the University of Utah as well as the cluster computing resources of the Max-Planck Institute for Astronomy Heidelberg. CC would also like to acknowledge contributions by Hans-Walter Rix, Rhys Seeburger, and Marten Scheuck. ACS, AB and CC acknowledge support from a \textit{Hubble Space Telescope} grant GO-16777.
AFK acknowledges funding from the Austrian Science Fund (FWF) [grant DOI 10.55776/ESP542].
M.A.C. acknowledges the support from Fondecyt Postdoctorado project No. 3230727.
\end{acknowledgments}

%

\vspace{5mm}
\facilities{\textit{HST}(STScI), MUSE@VLT}


\software{\texttt{IsoInterpFeh, IsoSplit, emcee}}



 \appendix
 \section{F625W Systematics}\label{appendix:F625W_systematics}
Here we show the results from age determinations using F625W in place of F606W (see Section~\ref{subsec:f625w}).   This includes results for both the same sample of stars as used in F606W, as well as for the larger sample that this filter combination enables due to the wider spatial coverage in F625W.  We note the photometric quality is much lower for individual stars in F625W relative to F606W.

\begin{figure}[h]
\centering
\includegraphics[width = \textwidth]{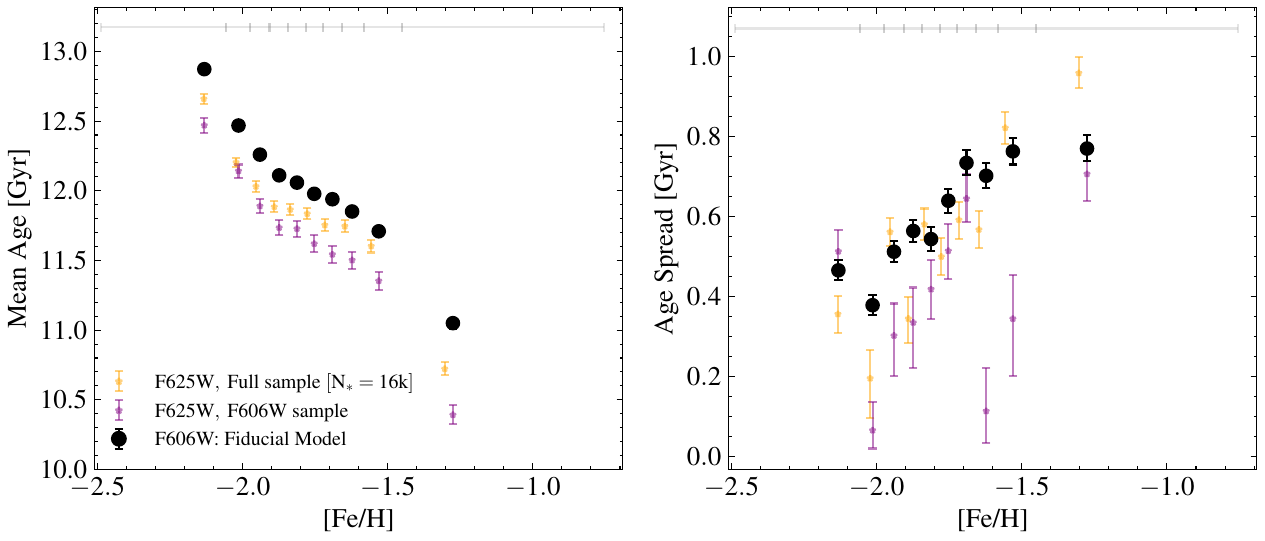}
\caption{\textbf{Mean Age vs. Age Spread Systematics for F625W:} Our mean age results are dependent on our choice of photometric band, though the age spread values are typically in good agreement regardless. See Section \ref{subsec:f625w}
 for more details.}\label{fig:f625w_systematics}
\end{figure}

 \section{Empirical Reddening Correction}\label{appendix:empirical_reddening_corr}

\omc's low Galactic latitude results in it having high foreground extinction. The \citet{Harris_1996} catalog (updated 2010) provides a reddening value for \omc\ of E(B-V)$=$0.12. However, using this value along with the Gaia-based distance \citep{Baumgardt_2021} results in a significant mismatch between our data and our isochrone models. This mismatch has been seen previously; for instance in \citet{Tailo_2016} they adopt an E(F435W-F625W) value of 0.242, translating to E(B-V)$=$0.157. \footnote{Note that throughout this section and paper we assume ${\rm A_V/E(B-V)} = 3.1$ and use the \textit{HST} ${\rm A_\lambda / A_V}$ values from http://stev.oapd.inaf.it/cgi-bin/cmd\_3.7}.

Apart from the isochrone mismatch, there is additional existing evidence for a higher extinction toward \omc\ than the \citet{Harris_1996} value.
\cite{Zhang_2023} use Gaia BP/RP spectra to forward model stellar atmospheric parameters, distances, and extinctions for over 200 million stars in our galaxy, with 1305 stars in \omc.  The extinction pattern of these stars suggests an increase in extinction toward the center of \omc. The median reddening for all \omc\ stars in their catalog is E(B-V)$=$0.139 while those shared with our catalog near the center of the cluster ($\sim$30 stars) have a median reddening of E(B-V)$=$0.191. In addition, \cite{Pancino_2024} derived differential extinction maps using photometry for 48 globular clusters and found an increase in reddening toward the center of \omc\ of up to 0.04 (see Figure~A.2 within).  These measurements motivate a re-determination of the E(B-V) value in \omc\ (specifically near the center where our data lie), incorporating the now better-known distance to the cluster.  

We derive an empirical correction for the reddening in \omc\ by comparing it to other Milky Way globular clusters.  We selected clusters with metallicity within 0.1 dex of \omc's dominant population ([Fe/H] $\sim -1.7$), available photometry in F606W and F814W from the ACS Globular Cluster Survey \citep[ACSGCS,][]{Sarajedini_2007}, and available distances from \citet{Baumgardt_2021}.  We use the two clusters with the lowest extinction: NGC 4147 and NGC 7089. These clusters have metallicities of -1.8 dex and -1.65 dex \citep{Harris_1996} respectively and have low reddening \citep[E(B-V)$=$0.02 and 0.06;][]{Harris_1996}. We note that the [$\alpha$/Fe] of these clusters also bracket the typical +0.3 value in \omc, with +0.38 in NGC4147 \citep{Villanova_2016} and +0.13 in NGC~7089 \citep{Recio-Blanco_2021}, and that the C+N+O enhancement matches the [$\alpha$/Fe] for the dominant [Fe/H]$= -1.7$ population in \omc \citep{Marino_2012}.  Therefore the stellar populations should be quite comparable to \omc, making them good candidates to use as reference populations for an empirical reddening correction. We adapt the procedure outlined in \citet{Marin-Franch_2009} to derive our new E(B-V) value.

\begin{figure}
\centering
\includegraphics[width = \textwidth]{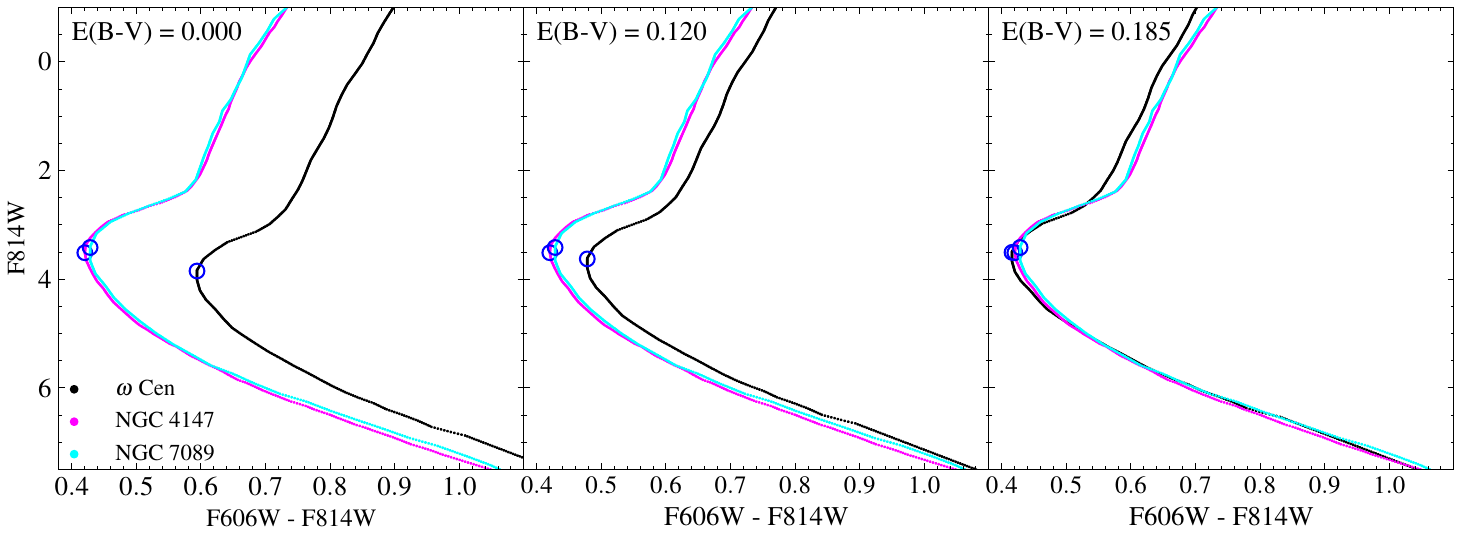}
\caption{Derivation of an empirical reddening correction for \omc.  All panels show the extinction-corrected ridgelines for two Milky Way Globular Clusters NGC~4147 and NGC~7089 (magenta and cyan lines).  The black line shows the dominant [Fe/H]$\sim -1.7$ metallicity population in \omc\ shifted by different extinctions in each panel.  The left panel shows the ridgeline with no extinction; the middle panel the ridgeline using the E(B-V) from \citet{Harris_1996} of 0.12, while the right panel shows our best-fit E(B-V)$=$0.185.}
\label{fig:empirical_ext_corr}
\end{figure}

For \omc{} stars we select those within $\pm$0.05 dex of the median metallicity of \omc\ (-1.75 \textless [Fe/H] \textless -1.65) for comparison with the other clusters. We then look at each cluster's color-magnitude-diagram using the absolute magnitudes based on distances in \citet{Baumgardt_2021} and draw by hand a ridgeline through the most-densely-populated regions from the RGB down through the MS.
After correcting the reference clusters ACSGCS photometry using the \citet{Harris_1996} reddening values we see that the two lie on top of one another well across the full CMD, confirming that the differences in metallicity will not contribute significantly to the reddening correction constraint. We then interpolate these points to ensure even and dense spacing.  We label the bluest point on the interpolated ridgeline the MSTO and then isolate a region on the MS with $ \rm +3 < m_{F814W}^{MSTO} < +1.5$ and a region on the RGB with $ \rm -2.5 < m_{F814W}^{MSTO} < -1.5$ to compare the clusters.  We then shift the \omc\ ridgeline with varying E(B-V).  For each E(B-V) value we calculate the sum of the absolute difference in color between NGC 4147 and \omc\ (NGC 5139) and again between NGC 7089 and \omc. We take the best-fit value to be the minimum of the sum of the absolute values for each region and then calculate the total color difference by taking the square-root of the sum of the squared color differences. By minimizing this combined color difference we find the best-fit value of E(B-V)$=$0.185. This process is shown in Fig.~\ref{fig:empirical_ext_corr}; all panels show NGC~4147 and NGC~7089 after correction for their \citet{Harris_1996} extinction values and each panel then shifts \omc's ridgeline based on varying extinction.  

Our derived value is independent of the assumed distance modulus since we are fitting only the colors in the region around the MSTO.  A potential significant difference may be due to helium variations in \omc, which are likely larger than in our comparison clusters and may result in a color shift relative to these clusters, which could result in a mismatch of our extinction.  To quantify its maximum contribution to the uncertainty we  apply a color shift to the data based on  the median color difference between the Y=0.245+1.5Z isochrone and the Y=0.40 isochrone in MS and RGB reference regions; the helium enriched isochrone is 0.047 and 0.042 mag bluer in these two regions respectively. These isochrones should represent the maximum range of helium abundances in cluster stars, and thus the difference between these two isochrones should encompass the potential error on the E(B-V).  Adding and subtracting this color difference we get differences of $\sim$0.04 in the inferred E(B-V) values, thus the uncertainty due to differing helium abundances should be smaller than this.

\section{\textbf{Provided Data Products}}
\label{appendix:provided_data_products}
The isochrone base model grids used for this analysis have been collected and described at https://doi.org/10.5281/zenodo.13810675.
\\
We provide a Machine-Readable-Table (MRT) with all of the age constraints used in our analysis for the base SGB sample ($N_* = 9129$). The information regarding each column is provided in Table \ref{table:mrt_ages}.

\begin{table}[h]
    \centering
    \begin{tabular}{l l l}
        \hline
         Column & Description & Unit \\ 
         \hline
         Star\_ID            &  oMEGACat Catalog Identifier & - \\
         Id\_MUSE            &  Nitschai et al. 2023 MUSE Catalog Identifier & - \\
         $\rm{[Fe/H]}$       &  Metallicity (from [M/H], assuming $ \rm{[\alpha/Fe]}$ = 0.3, with atomic diffusion correction) & dex \\
 
         age\_fiducial       &  Stellar ages using fiducial model parameters & Gyr\\
         age\_fiducial\_err  &  Stellar age uncertainties using fiducial model parameters & Gyr\\

         age\_cno\_fixed     &  Stellar ages when using CNO-Fixed isochrone & Gyr\\
         age\_cno\_fixed\_err  &  Stellar age uncertainties when using CNO-Fixed isochrones & Gyr\\

         age\_afep2          &  Stellar ages when setting [$\rm{\alpha/Fe}$] = 0.2 & Gyr\\
         age\_afep2\_err      &  Stellar age uncertainties when setting [$\rm{\alpha/Fe}$] = 0.2 & Gyr\\

         age\_F625W            &  Stellar ages when using F625W instead of F606W & Gyr\\
         age\_F625W\_err        &  Stellar age uncertainties when using F625W instead of F606W & Gyr\\

         age\_y40            &  Stellar ages when setting Y=0.40 & Gyr\\
         age\_y40\_err        &  Stellar age uncertainties when setting Y=0.40 & Gyr\\

         age\_dist\_sys       &  Stellar ages when subtracting distance uncertainty (0.20 kpc) & Gyr\\
         age\_dist\_sys\_err   &  Stellar age uncertainty when subtracting distance uncertainty & Gyr\\

         age\_feh\_offset\_sys       &  Stellar ages when adding $\rm{[Fe/H]}$ offset (0.064) & Gyr\\
         age\_feh\_offset\_sys\_err   &  Stellar age uncertainties when adding $\rm{[Fe/H]}$ offset & Gyr\\
        \hline
    \end{tabular}
    \caption{Ages and Age Uncertainties: Ages (and their errors) for the full SGB sample as well as each systematic contstraint.}
    \label{table:mrt_ages}
\end{table}

\bibliography{main_bib}
\bibliographystyle{aasjournal}



\end{document}